\documentclass[12pt]{article}
\usepackage[utf8]{inputenc}
\usepackage[english]{babel}
\usepackage{a4wide}
\usepackage{amsfonts}
\usepackage{amssymb}
\usepackage{graphics,amsmath}
\usepackage{graphicx}
\usepackage{cancel}
\usepackage{cite}
\usepackage{mathrsfs}
\usepackage[usenames,dvipsnames,svgnames,table]{xcolor}
\usepackage{slashed}
\usepackage{etoolbox}

\usepackage[font=small,labelfont=bf]{caption}

\usepackage{tikz}
\usetikzlibrary{decorations.markings}

\tikzset{test/.style n args={3}{
    postaction={
    decorate,
    decoration={
    markings,
    mark=between positions 0 and \pgfdecoratedpathlength step 0.5pt with {
    \pgfmathsetmacro\myval{multiply(
        divide(
        \pgfkeysvalueof{/pgf/decoration/mark info/distance from start}, \pgfdecoratedpathlength
        ),
        100
    )};
    \pgfsetfillcolor{#3!\myval!#2};
    \pgfpathcircle{\pgfpointorigin}{#1};
    \pgfusepath{fill};}
}}}}

\def\hybrid{\topmargin -20pt    \oddsidemargin 0pt
        \headheight 0pt \headsep 0pt
        \textwidth 6.35in       
        \textheight 9.25in       
        \marginparwidth .875in
        \parskip 5pt plus 1pt   \jot = 1.5ex}

\hybrid

\def\baselinestretch{1.2}

\catcode`\@=11

\def\marginnote#1{}
\newcount\hour
\newcount\minute
\newtoks\amorpm
\hour=\time\divide\hour by60
\minute=\time{\multiply\hour by60 \global\advance\minute by-\hour}
\edef\standardtime{{\ifnum\hour<12 \global\amorpm={am}%
        \else\global\amorpm={pm}\advance\hour by-12 \fi
        \ifnum\hour=0 \hour=12 \fi
        \number\hour:\ifnum\minute<10 0\fi\number\minute\the\amorpm}}
\edef\militarytime{\number\hour:\ifnum\minute<10 0\fi\number\minute}

\def\draftlabel#1{{\@bsphack\if@filesw {\let\thepage\relax
   \xdef\@gtempa{\write\@auxout{\string
      \newlabel{#1}{{\@currentlabel}{\thepage}}}}}\@gtempa
   \if@nobreak \ifvmode\nobreak\fi\fi\fi\@esphack}
        \gdef\@eqnlabel{#1}}
\def\@eqnlabel{}
\def\@vacuum{}
\def\draftmarginnote#1{\marginpar{\raggedright\scriptsize\tt#1}}

\def\draft{\oddsidemargin -.5truein
        \def\@oddfoot{\sl preliminary draft \hfil
        \rm\thepage\hfil\sl\today\quad\militarytime}
        \let\@evenfoot\@oddfoot \overfullrule 3pt
        \let\label=\draftlabel
        \let\marginnote=\draftmarginnote
   \def\@eqnnum{(\theequation)\rlap{\kern\marginparsep\tt\@eqnlabel}%
\global\let\@eqnlabel\@vacuum}  }

\def\preprint{\twocolumn\sloppy\flushbottom\parindent 2em
        \leftmargini 2em\leftmarginv .5em\leftmarginvi .5em
        \oddsidemargin -.5in    \evensidemargin -.5in
        \columnsep .4in \footheight 0pt
        \textwidth 10.in        \topmargin  -.4in
        \headheight 12pt \topskip .4in
        \textheight 6.9in \footskip 0pt
        \def\@oddhead{\thepage\hfil\addtocounter{page}{1}\thepage}
        \let\@evenhead\@oddhead \def\@oddfoot{} \def\@evenfoot{} }

\def\numberbysection{\@addtoreset{equation}{section}
        \def\theequation{\thesection.\arabic{equation}}}

\def\underline#1{\relax\ifmmode\@@underline#1\else
        $\@@underline{\hbox{#1}}$\relax\fi}

\def\titlepage{\@restonecolfalse\if@twocolumn\@restonecoltrue\onecolumn
     \else \newpage \fi \thispagestyle{empty}\c@page\z@
        \def\thefootnote{\fnsymbol{footnote}} }

\def\endtitlepage{\if@restonecol\twocolumn \else \newpage \fi
        \def\thefootnote{\arabic{footnote}}
        \setcounter{footnote}{0}}  

\catcode`@=12
\relax

\def\figcap{\section*{Figure Captions\markboth
        {FIGURECAPTIONS}{FIGURECAPTIONS}}\list
        {Figure \arabic{enumi}:\hfill}{\settowidth\labelwidth{Figure
999:}
        \leftmargin\labelwidth
        \advance\leftmargin\labelsep\usecounter{enumi}}}
 \relax
\def\tablecap{\section*{Table Captions\markboth
        {TABLECAPTIONS}{TABLECAPTIONS}}\list
        {Table \arabic{enumi}:\hfill}{\settowidth\labelwidth{Table
999:}
        \leftmargin\labelwidth
        \advance\leftmargin\labelsep\usecounter{enumi}}}
 \relax
\def\reflist{\section*{References\markboth
        {REFLIST}{REFLIST}}\list
        {[\arabic{enumi}]\hfill}{\settowidth\labelwidth{[999]}
        \leftmargin\labelwidth
        \advance\leftmargin\labelsep\usecounter{enumi}}}
 \relax
\makeatletter
\newcounter{pubctr}
\def\publist{\@ifnextchar[{\@publist}{\@@publist}}
\def\@publist[#1]{\list
        {[\arabic{pubctr}]\hfill}{\settowidth\labelwidth{[999]}
        \leftmargin\labelwidth
        \advance\leftmargin\labelsep
        \@nmbrlisttrue\def\@listctr{pubctr}
        \setcounter{pubctr}{#1}\addtocounter{pubctr}{-1}}}
\def\@@publist{\list
        {[\arabic{pubctr}]\hfill}{\settowidth\labelwidth{[999]}
        \leftmargin\labelwidth
        \advance\leftmargin\labelsep
        \@nmbrlisttrue\def\@listctr{pubctr}}}
 \relax
\makeatother
\newskip\humongous \humongous=0pt plus 1000pt minus 1000pt

\newif\ifdtup

\relax

\def\be{\begin{equation}}
\def\ee{\end{equation}}
\def\ba{\begin{eqnarray}}
\def\ea{\end{eqnarray}}

\def\u { \{ u\}}

\def\no{\noindent}

\def\IR{\relax{\rm I\kern-.18em R}}
\def\II{\relax{\rm 1\kern-.35em1}}

\renewcommand{\theequation}{\thesection.\arabic{equation}}
\csname @addtoreset\endcsname{equation}{section}

\def\IR{\relax{\rm I\kern-.18em R}}
\def\inv{^{\raise.15ex\hbox{${\scriptscriptstyle -}$}\kern-.05em 1}}

\newcommand{\bra}[1]{\left\langle #1 \right|}
\newcommand{\ket}[1]{\left| #1 \right\rangle}

\let\emptyset\varnothing

\begin{document}

\begin{titlepage}
\begin{center}

\vskip .5in

{\LARGE Cutting the cylinder into squares: The square form factor}
\vskip 0.4in

{\bf Juan Miguel Nieto}
\vskip 0.1in

Departamento de F\'{\i}sica Te\'orica \\
Universidad Complutense de Madrid \\
$28040$ Madrid, Spain \\
{\footnotesize{\tt juanieto@ucm.es}}

\end{center}

\vskip .4in

\centerline{\bf Abstract}
\vskip .1in
\no

In this article we present a method for constructing two-point functions in the spirit of the hexagon proposal, which leads us to propose a ``square form factor''. Since cutting the square gives us two squares, we can write a consistency condition that heavily constrains such form factors. In particular, we are able to use this constraint to reconstruct the Gaudin through the forest expansion of the determinant appearing in its definition. We also use this procedure to compute the norm of off-shell Bethe states for some simple cases.

\noindent

\vskip .4in
\noindent

\end{titlepage}

\vfill
\eject

\def\baselinestretch{1.2}


\baselineskip 20pt


\section{Introduction}

A consistent method to define a quantum field theory is to construct observables starting from form factors. Form factors are matrix elements of local observables computed in the basis of asymptotic ingoing and outgoing states. Such matrix elements fulfil the \emph{Smirnov's Axioms} \cite{Smirnov}, a set of restrictions on their analytical properties. This construction is non-perturbative and restricts the reconstructed theory to be local and a Wightman field theory.

This construction has been successfully applied to different systems like the Sine-Gordon model \cite{Babujian_1999,Babujian_2002}, the Thirring model \cite{Nakayashiki_2002,Takeyama_2003} and the SU(N) Lieb-Lininger model \cite{Babujian_2006}. Concerning the $AdS_5/CFT_4$ correspondence, we can find the first steps towards understanding form factors in \cite{WSFF1,WSFF2}, where the authors focus on the spin chain description and the near-plane-wave limit of the string description.

Regarding direct computations of correlation functions, the Quantum Spectral Curve has already solved (at least formally) the spectral problem in $\mathcal{N}=4$ SYM \cite{GKLV1,GFS1,GKLV2,CV1,CV2,GFS2,KLV,CV3,CV4,CV5} and thus the computation of two-point functions due to conformal symmetry. As a consequence, the focus has partially shifted to three-point functions. Although neither the interest nor the use of integrability are new in this context (see, for example, \cite{3pf1,3pf2,vertex1,3pf3}), recently a very simple and successful proposal for an all-loop computation of three-point functions based on cutting the pair of pants into two separated hexagons has been put forward \cite{BKV,Hex2,Hex3}. Interestingly, these hexagons can be understood as a form factor involving a defect operator. Since the operator involved is non-local, these form factor fulfil a slightly modified version Smirnov's axioms \cite{nonlocalFF}. It has been noticed that the hexagon form factors alone are enough to construct higher point functions \cite{Hexagonalization1,Hexagonalization2,Hexagonalization3,Hexagonalization4} due to the feasibility of obtaining form factors with a larger number of sides just by gluing hexagons.

\begin{figure}[t]
\begin{center}

\begin{tikzpicture}[scale=1.8]
\draw (0,-0.5) -- (0,1.5);
\draw (1,-0.5) -- (1,1.5);
\draw (0.5,1.5) ellipse (0.5 and 0.2);
\draw [densely dashed] (1,-0.5) arc (0:180:0.5 and 0.2);
\draw (0,-0.5) arc (180:360:0.5 and 0.2);
\draw [fill] (0.5,1.3) circle [radius=0.04];
\draw [fill] (0.5,-0.7) circle [radius=0.04];
\draw [thick] (0.5,1.3) -- (0.5,-0.7);

\draw [ultra thick] (1.3333333,0.4) -- (1.6666666,0.4);
\draw [ultra thick] (1.3333333,0.6) -- (1.6666666,0.6);

\draw [densely dashed] (2,0) --(2,1);
\draw (2, 0) -- (3,0);
\draw (2, 1) -- (3,1);
\draw [densely dashed] (3,0) --(3,1);
\draw [fill] (2.5,1) circle [radius=0.04];
\draw [fill] (2.5,0) circle [radius=0.04];
\draw [thick] (2.5,0) -- (2.5,1);
\draw [ultra thick] (3.3333333,0.5) -- (3.6666666,0.5);
\draw [ultra thick] (3.5,0.3333333) -- (3.5,0.66666666);
\draw [densely dashed] (4,0) --(4,1);
\draw (4, 0) -- (5,0);
\draw (4, 1) -- (5,1);
\draw [densely dashed] (5,0) --(5,1);
\draw [fill] (4.5,1) circle [radius=0.04];
\draw [fill] (4.5,0) circle [radius=0.04];
\draw [thick] (4.5,0) -- (4.5,1);
\draw [fill] (3.96,0.46) rectangle (4.04,0.54);
\draw [fill] (4.96,0.46) rectangle (5.04,0.54);
\draw [dotted,thick] (4,0.5) -- (5,0.5);

\draw [ultra thick] (5.3333333,0.5) -- (5.6666666,0.5);
\draw [ultra thick] (5.5,0.3333333) -- (5.5,0.66666666);

\draw [densely dashed] (6,0) --(6,1);
\draw (6, 0) -- (7,0);
\draw (6, 1) -- (7,1);
\draw [densely dashed] (7,0) --(7,1);
\draw [fill] (6.5,1) circle [radius=0.04];
\draw [fill] (6.5,0) circle [radius=0.04];
\draw [thick] (6.5,0) -- (6.5,1);
\draw [fill] (5.96,0.293333333) rectangle (6.04,0.373333333);
\draw [fill] (6.96,0.293333333) rectangle (7.04,0.373333333);
\draw [fill] (5.96,0.626666666666) rectangle (6.04,0.706666666666);
\draw [fill] (6.96,0.626666666666) rectangle (7.04,0.706666666666);
\draw [dotted,thick] (6,0.333333333) -- (7,0.33333333333333);
\draw [dotted,thick] (6,0.666666666) -- (7,0.66666666666666);

\draw [ultra thick] (7.3333333,0.5) -- (7.6666666,0.5);
\draw [ultra thick] (7.5,0.3333333) -- (7.5,0.66666666);
\draw [fill] (7.9,0.5) circle [radius=0.03];
\draw [fill] (8.1,0.5) circle [radius=0.03];
\draw [fill] (8.3,0.5) circle [radius=0.03];
\end{tikzpicture}
\caption{Pictorial representation of the cutting process applied to the cylinder. Note that cutting the cylinder just once imposes both mirror sides to have the same content.} \label{cuttingprocedure}
\end{center}
\end{figure}
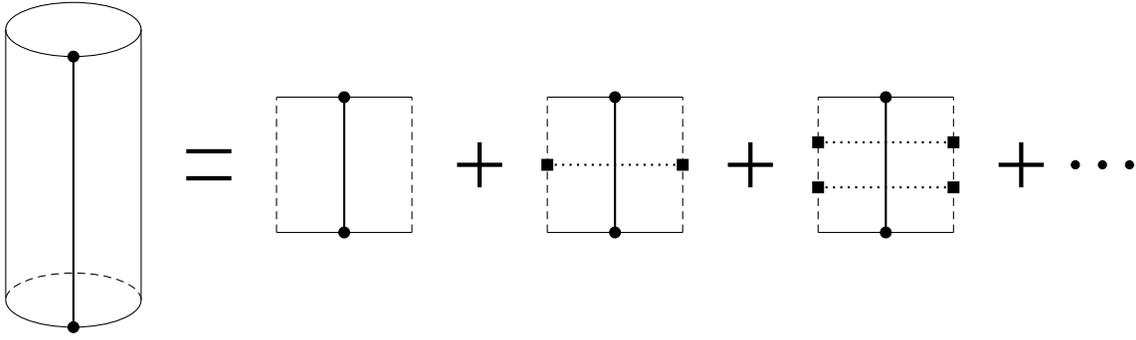

In this article we argue that, together with the hexagon form factor, we can define a square ``form factor'', more akin to usual form factors, related to the cutting of the cylinder (see figure~\ref{cuttingprocedure}). We write form factor between quotation marks because it would correspond to the form factor of the identity, which is sometimes not considered truly a form factor but still satisfy Smirnov's axioms. However, here we will construct the square starting from states constructed using the Algebraic Bethe Ansatz instead of states constructed with Zamolodchikov-Faddeev (ZF) operators, hence it does not fulfil the usual Smirnov's axioms but a different ones. Despite their differences there exists a relationship between both squares, we will postpone the analysis of ZF squares to future works.

One of the most important differences between the square and the hexagon form factor refers to the processes of gluing and cutting: when we glue together two squares we get again a square and vice-versa when cutting. We can exploit this feature to write down restrictions on the analytic properties of squares and reconstruct the scalar product of states. In particular, we will explicitly reconstruct the Gaudin determinant, as we are going to focus mainly on a system with $SU(2)$ symmetry.

Another interesting characteristic of the square is how mirror excitations appear in the gluing procedure. In contrast with the hexagon form factor, the $\mathcal{N}=4$ SYM square ``form factor'' is invariant under the full $PSU(2|2)^2_{\text{c.e.}}$ symmetry. As the central extension includes the dilatation operator, some configurations have to vanish and, in particular, the number of excitations in the upper and lower edges has to be the same when no mirror excitations are present. Thus, the gluing of two squares should involve only a finite sum due to symmetry instead of the infinite sum that appears when gluing of two hexagons. This makes the square a simpler setting where we can get a better understanding of this procedure.

It is worth pointing that a different square form factor from the one proposed here has appeared in the context of correlation functions of three separated operators connected by Wilson lines \cite{HexagonWL}. Despite the different nature of both squares, we expect them to fulfil similar consistency conditions.

The article is structured as follows. In section 2 we review some concepts related to the Algebraic Bethe Ansatz. In section 3 we present the consistency condition for the square and use it to reconstruct the normalization of on-shell Bethe states of the $SU(2)$ spin chain. In section 4 we extend the analysis to compute the normalization of off-shell Bethe states for the case of two and three excitations. We also compute the three highest orders in the large length expansion of the normalization of general $SU(2)$ off-shell states. In section 5 we close with summary, conclusions and comments on some points that should be clarified in further works. In appendix A we review Kirchhoff's matrix-tree theorem and its application to the Gaudin determinant. In appendix B we comment a little more on the form of the consistency condition for $SU(N)$ spin chains. In appendix C we explain how to recover the space-time dependence of the two point function. In appendix D we expand on the constraints imposed by crossing symmetry. In appendix E we sketch the computation of the leading mirror corrections to the squares with zero and one physical excitation.

\section{The $SU(2)$ Algebraic Bethe Ansatz}

In this section we present the basis of the Algebraic Bethe Ansatz (ABA) for $SU(2)$ models in order to fix the notation we will use through this article. We are going to work with the rational 6-vertex model R-matrix
\begin{align}
	R_{1,2} (\lambda) &=\alpha (u,v) \sum_{a=1}^2 E_{aa} \otimes E_{aa} +\beta(u,v) \sum_{1\leq a\neq b \leq 2} E_{aa} \otimes E_{bb} + \sum_{1\leq a\neq b \leq 2} \gamma_{ab} (u,v) E_{ab} \otimes E_{ba} \notag \\
	&=\lambda (\mathbb{I}_a \otimes \mathbb{I}_b ) +\frac{i\eta }{2} \sum_j{(\sigma^j_a \otimes \sigma^j_b)} \ ,
\end{align}
but our results can be easily generalized to the trigonometric 6-vertex model by an appropriate substitution of the functions $\alpha$, $\beta$ and $\gamma$. Here $E_{ab}$ is the matrix with zeros in every position except the $a,b$ position, which contains a $1$.

The monodromy matrix, written as a matrix on the auxiliary space, has the following structure
\begin{equation}
	T_a (\lambda )=\overrightarrow{\prod_{j=1}^L} \mathcal{L}_{a,j}(u)=\mathcal{L}_{a,1}(u) \mathcal{L}_{a,2}(u) \dots \mathcal{L}_{a,L}(u) =\left( \begin{array}{cc}
	A(\lambda ) & B(\lambda ) \\
	C(\lambda ) & D(\lambda )
	\end{array}\right) \ , \label{Monodromy}
\end{equation}
where $\mathcal{L}_{a,j}(u)$ is the Lax matrix associated to the specific model. The action of these four operators over the reference state is given by
\begin{align}
	A(\lambda ) \ket{0} &= a(\lambda ) \ket{0} \ , & B(\lambda ) \ket{0} &\neq 0 \ , \\
	D(\lambda ) \ket{0} &= d(\lambda ) \ket{0} \ , & C(\lambda ) \ket{0} &= 0  \ ,
\end{align}
while their commutation relations are fixed by the RTT relation
\begin{equation}
	R_{1,2} (\lambda - \mu ) T_1 (\lambda ) T_2 (\mu) = T_2 (\mu) T_1 (\lambda) R_{1,2} (\lambda - \mu ) \label{RTT} \ .
\end{equation}
Of the relations codified in this equation, we will only need the following seven of them
\begin{align}
	&[B(\lambda ) , B(\mu )]=[C(\lambda ) , C(\mu )]=0 \ , \label{commBB}\\
	&A(\lambda ) B(\mu )=f(\mu , \lambda) B(\mu ) A(\lambda ) +g(\lambda , \mu ) B(\lambda ) A(\mu ) \ ,\label{commAB}\\
	&D(\lambda ) B(\mu )=f(\lambda, \mu) B(\mu ) D(\lambda ) + g(\mu , \lambda ) B(\lambda ) D(\mu ) \ , \label{commDB} \\
	&C(\mu ) A(\lambda )=f(\mu , \lambda) A(\lambda ) C(\mu ) +g(\lambda , \mu ) A(\mu ) C(\lambda ) \ ,\label{commCA}\\
	&C(\mu ) D(\lambda )=f(\lambda, \mu) D(\lambda ) C(\mu ) + g(\mu , \lambda ) D(\mu ) C(\lambda ) \ , \label{commCD}\\
	&[C(\lambda ) , B(\mu) ]=g(\lambda , \mu ) \left[ A(\lambda ) D(\mu ) - A(\mu ) D(\lambda ) \right] \ , \label{commCB}
\end{align}
where, for convenience, we have introduced the functions \footnote{Using these functions instead of their explicit expressions allow us to generalize our results to the trigonometric model just by substituting their expression by the corresponding one without altering any formula. We also want to remark that we are defining $f(u,v)=1+g(u,v)$ in contrast with e.g. \cite{Faddeev}, where $f(u,v)=1-g(u,v)$.}
\begin{align*}
	f(\lambda , \mu) &=\frac{\alpha (\lambda , \mu)}{\beta(\lambda , \mu)} \ , & f(\lambda , \mu) &=\frac{\lambda - \mu +i}{\lambda - \mu} \ , & f(\mu , \lambda) &=\frac{\lambda - \mu -i}{\lambda - \mu} \ , \\
	g(\lambda , \mu )&=\frac{\gamma_{12} (\lambda , \mu)}{\beta(\lambda , \mu)}\ , & g(\lambda , \mu )&=\frac{i}{\lambda - \mu} \ , & g(\mu , \lambda )&=\frac{-i}{\lambda - \mu} \ .
\end{align*}

Since the trace of the monodromy matrix, called transfer matrix, commute with itself for different values of the spectral parameter $\lambda$, its power expansion produces $L-1$ commuting operators, any of which can be chosen as Hamiltonian. We can construct the Hilbert space by applying a stack of $B$ operators over the reference state, but such space is overcomplete as not all the states created by this process are eigenstates of the transfer matrix. The states that are not eigenstates of the transfer matrix are not physical, hence we will call them \emph{off-shell Bethe states}. Only the states constructed with B operators with some precise sets of rapidities are eigenstates of the transfer matrix and hence true physical states. We will call such states \emph{on-shell Bethe states}. The admissible sets of rapidities are those that fulfil the Bethe Ansatz Equations
\begin{equation}
	\frac{a (u_j)}{d(u_j)}=\prod_{k\neq j} \frac{f(u_j,u_k)}{f(u_k,u_j)} \ . \label{BAE}
\end{equation}

As most of the time we will work with sets of rapidities, in the remainder of the article the following shorthand notation will prove useful as it reduces the amount of products on the formulas
\begin{align*}
	f( \{\bar{\alpha}\},\{\alpha\}) &=\prod_{\substack{u_i\in \alpha \\ v_j \in \bar{\alpha}}} f(v_j,u_i) \ , & f^{\neq} (\{\alpha \} , \{ \alpha \} ) &=\prod_{\substack{u_i,u_j\in \alpha \\ u_i\neq u_j}} f(u_i , u_j) \ , & a (\{\bar{\alpha}\}) &=\prod_{v_j \in \bar{\alpha}} a (v_j) \ .
\end{align*}

\section{From the Gaudin determinant to the square and back}

In this section we propose the \emph{square form factor} and a consistency condition for it. After proposing this consistency equation, we verify that the Gaudin norm fulfils it as a check for our proposal. Reversing the argument, we are able to reconstruct the Gaudin determinant from the consistency condition and some initial conditions and analytic considerations.

\subsection{The square bootstrap and the Gaudin determinant}

Inspired by the BKV idea of cutting the three-point correlation function into two hexagon form factors \cite{BKV}, we apply the same procedure to two-point functions, i.e. a cylinder worldsheet. In this case we will obtain two squares after cutting two times, each with two physical sides and two mirror sides, being the last the new sides created by this process. We decided to define our square by cutting two times the cylinder instead of one time because in the later case the two mirror sides of the square are constrained to have the same number of excitations (see figure~\ref{cuttingprocedure}).

The most general square is characterized by four sets of rapidities, one for each of the physical sides and one for each of the mirror sides. For the most part of this article the two mirror sides will be empty, so we only have to indicate the content of the two physical sides. Therefore we denote the square by $S_L ( \u \rightarrow \{ w \})$, where the subindex indicates the length of the physical sides. If both sets of rapidities are equal, we will just write $S_L ( \u )=S_L ( \u \rightarrow \{ u \})$ to alleviate notation. With this notation the relationship between the scalar product of Bethe states and the square is given by
\begin{equation}
	\Big\langle 0 \Big| \prod_i C(v_i) \prod_j B(u_j) \Big| 0 \Big\rangle =S_L (\{ u \} \rightarrow \{ v \} ) + \text{ mirror corrections} \ .
\end{equation}
The superposition of two Bethe states gives us the two-point function of an operator at zero with an operator at infinity, i.e., the normalization of the two-point functions. Despite this being an unphysical quantity, it is important to compute it to correctly normalize our states. In this article we will consider the $L\gg 1$ limit, so we can ignore mirror corrections. In appendix~\ref{mirrorappendix} we will comment about how to include mirror corrections.

Although the construction of the hexagon form factor from symmetry and crossing arguments is very simple and elegant, a similar construction for the square presents some obstructions. Since here we are addressing the problem from the weak coupling limit, we do not have access to the crossing transformation. This transformation entails shifting the rapidity of an excitation in such a way that it crosses some particular cuts in the rapidity plane. However these cuts have a width of $4g$, thus they close at weak coupling and the mirror and crossing transformations are inaccessible perturbatively \cite{BetheGKP}. Nevertheless, an all-loop construction of the square should behave nicely under crossing transformations, e.g. we expect that $S_L ( \{ u^{\text{cross}} \} \rightarrow \{ w^{\text{cross}} \})=S_L ( \{ w \} \rightarrow \{ u \})$.

Among the several differences that exist between this square and the hexagon form factor, we are interested in the one concerning the cutting and the gluing of form factors. The cutting operation applied to the square remains well defined after we employ it once, \footnote{Although this is true for cutting the square vertically, horizontal cuts should behave differently. We will not address that kind of cuts in this article.} which means that we can continue cutting a square into smaller ones. We can reverse the argument and state that gluing two squares reconstructs one square. This idea can be used as a ``bootstrap procedure'' to construct such square. If we consider only the asymptotic part of the square (which means directly gluing them without adding excitations to the mirror edges), the reconstruction formula will be given by
\begin{equation}
	S_L ( \{ u \} \rightarrow \{ v \} )=\sum_{\substack{\alpha \cup \bar{\alpha} =\{ u\} \\ \beta \cup \bar{\beta} =\{ v\}}} w (\{ \alpha \} , \{ \bar{\alpha} \} ,\{ \beta \} , \{ \bar{\beta} \} ) S_{l_1} ( \{ \alpha \} \rightarrow \{ \beta \} ) S_{l_2} ( \{ \bar{\alpha} \} \rightarrow \{ \bar{\beta} \} ) \ , \label{bootstrap}
\end{equation}
where the weight $w (\{ \alpha \} , \{ \bar{\alpha} \} ,\{ \beta \} , \{ \bar{\beta} \} )$, which we will call \emph{breaking factor}, takes into account the cutting of the physical states into two. We have also defined $L=l_1+l_2$. In this article we will work in the $L \sim l_1 \sim l_2 \gg 1$, so we will not consider any potential problems that could arise from the discrete nature of the states when we cut them. As we said before, this limit also allow us to ignore the mirror corrections as they are exponentially suppressed.

To construct the breaking factors we proceed in a similar manner as \cite{composite}, i.e. we write the monodromy matrix as the product of two monodromy matrices, each associated to a subchain of the original chain. Using the definition of the monodromy matrix given in equation~(\ref{Monodromy}) we can rewrite the product of $L$ Lax matrices into two separate products with $l_1$ and $l_2$ Lax matrices each. Equating the entries of the auxiliary space we can prove that
\begin{equation}
	T^{(L)}_a(u)=T^{(l_1)} (u) \otimes T^{(l_2)} (u) \Longrightarrow \left\{ \begin{array}{c}
	B_L (u) = B_{l_1} (u) \otimes D_{l_2} (u) + A_{l_1} (u) \otimes B_{l_2} (u) \\
	C_L (u) = C_{l_1} (u) \otimes A_{l_2} (u) + D_{l_1} (u) \otimes C_{l_2} (u) 
\end{array}	 \right. \ . 
\end{equation}
These formulas provide us with the recipe to break the creation and annihilation operators for the case of one excitation. For the general square we need the expression of several creation and annihilation operators. As the RTT relations involve at most two rapidities, it is easy to argue that the structure of the weight functions can be fixed from the breaking factor for two excitations. For the case of creation operators we have
\begin{align*}
	B(u) B(v) &= B_{l_1}(u) B_{l_1}(v) \otimes D_{l_2}(u) D_{l_2}(v) + B_{l_1}(u) A_{l_1}(v) \otimes D_{l_2}(u) B_{l_2}(v) \notag \\
		&+ A_{l_1}(u) B_{l_1}(v) \otimes B_{l_2}(u) D_{l_2}(v) + A_{l_1}(u) A_{l_1}(v) \otimes B_{l_2}(u) B_{l_2}(v) = \notag \\
		&=B_{l_1}(u) B_{l_1}(v) \otimes D_{l_2}(u) D_{l_2}(v) + f(u,v) B_{l_1}(u) A_{l_1}(v) \otimes B_{l_2}(v) D_{l_2}(u) \notag \\
		&- g(u,v)  B_{l_1}(u) A_{l_1}(v) \otimes B_{l_2}(u) D_{l_2}(v)  + f(v,u) B_{l_1}(v) A_{l_1}(u) \otimes B_{l_2}(u) D_{l_2}(v) \notag \\
		&+ g(u,v) B_{l_1}(u) A_{l_1}(v) \otimes B_{l_2}(u) D_{l_2}(v) + A_{l_1}(u) A_{l_1}(v) \otimes B_{l_2}(u) B_{l_2}(v) \ .
\end{align*}
In the second equality we have used RTT equations (\ref{commAB}) and (\ref{commDB}) to move all $A$ and $D$ operators to the right of the $B$ operators. Applying this expression to the reference state gives us the contribution from $B$ operators to the weight in the case of two excitations. As argued before, the case of more excitations is just an extension of this computation and gives us an equivalent structure. A similar procedure can be performed for annihilation operators by instead moving the $A$ and $D$ operators to the left of the $C$ operators and applying the expression to the dual reference state. The final expression for the breaking factor is the product of the weights associated to $B$ and $C$ operators
\begin{align}
	w (\{ \alpha \} , \{ \bar{\alpha} \} ,\{ \beta \} , \{ \bar{\beta} \} ) &= w_B (\{ \alpha \} , \{ \bar{\alpha} \} ) w_C (\{ \beta \} , \{ \bar{\beta} \} ) \label{weights} \\
	w_B (\{ \alpha \} , \{ \bar{\alpha} \} ) &= f(\alpha , \bar{\alpha}) a_{l_1} (\bar{\alpha}) d_{l_2} (\alpha) \ , \notag \\
	w_C (\{ \beta \} , \{ \bar{\beta} \} ) &= f(\bar{\beta} , \beta) d_{l_1} (\bar{\beta}) a_{l_2} (\beta) \ . \notag
\end{align}
Despite the simplicity of the result, this computation becomes rather messy when we move to $SU(N)$ for $N>2$. A small discussion about it has been collected in appendix~\ref{BFappendix}, together with the computation of the breaking factor for $SU(2)$ when we work with operators that satisfy the Zamolodchikov-Faddeev algebra.

Let us perform now a first check of our proposal. The Gaudin determinant provides us the normalization of states of the XXX spin chain written using the Algebraic Bethe Ansatz. Its explicit expression is
\begin{align}
	G(\{u\}) &=i^N  f^{\neq} (\u , \u) \det \Phi \ , & \Phi_{ab} &=-\frac{\partial}{\partial u_b} \left( \frac{a (u_a)}{d (u_a)} \prod_{k\neq a} \frac{f(u_c , u_a)}{f(u_a , u_c)} \right) \ . \label{gaudin}
\end{align}
The proof that such determinant fulfils the consistency condition we have proposed can already be found in the literature, see for example eq. (36) in \cite{diagonalFF}. Instead, for later convenience it proves better to use the Kirchhoff's matrix-tree theorem \cite{Kirchhoff} to rewrite the determinant expression as a sum over graphs. We have summarized this construction in appendix~\ref{Kirchhoffappendix} and we refer to \cite{Tree} and references therein for more details. In this forest expansion all the dependence on the length of the chain appears only through the factors associated to the root nodes of the trees. If we break the length $L$ into $l_1+l_2$, we have to add an extra sum over trees having either $l_1 p'$ or $l_2 p'$ as root factor. Grouping terms by root nodes and rapidity labels involved in each set of trees allows us to reconstruct the two factors $S_{l_1} ( \{ \alpha \} )$ and $S_{l_2} ( \{ \bar{\alpha} \} )$, with the corresponding sum coming from the forest expansion and choices of root nodes becoming the sum of equation (\ref{Monodromy}). The appearance of the breaking factor comes not from separating the determinant into trees, but from separating the prefactor $f^{\neq} (\u , \u)$.


It might seem strange that the Gaudin determinant directly fulfils this relation because the square should be an off-shell construction while the states involved in the Gaudin determinant are on-shell Bethe states. This is so because the terms dependent on the Bethe Ansatz Equations appear as
\begin{equation}
	S_L (\u)= \{ \text{On-shell part}\} + \{ \text{Bethe eq.} \} \cdot \{ \text{Off-shell contributions} \} \ ,
\end{equation}
so the on-shell part should also fulfils the same consistency equation. In the next section we will check for some simple examples that the square has this structure.

\subsection{Reconstructing the Gaudin determinant: some particular cases} \label{gaudinreconstruction}

After checking that the Gaudin determinant fulfils the consistency condition~(\ref{bootstrap}), we are going to try to reconstruct it from the consistency condition. First we present the reconstruction for the particular cases from zero up to three excitations in order to clarify the procedure. After that we will provide the fully reconstructed expression for a general number of excitations, up to some functions that cannot be fixed just using the consistency condition. We will fix them using comparison with explicit ABA computation and analytic considerations.

Basic idea of this reconstruction is to use the length structure in the consistency condition to impose constraints. While $S_{l_1}$ and $S_{l_2}$ depend explicitly on $l_1$ and $l_2$ respectively and the weight factors depend on those lengths implicitly though the momentum factors $a$ and $d$, $S_L$ cannot depend on these lengths but on the form $(l_1+l_2)$.

Apart from the consistency condition, we will need two additional properties of the square to fix it. In particular we will need that
\begin{itemize}
	\item The square is symmetric under the exchange of two rapidities, i. e.
	\begin{equation}
		S_L (u_1 , \dots , u_i , \dots , u_j , \dots )=S_L (u_1 , \dots , u_j , \dots , u_i , \dots ) \ .
	\end{equation}
	\item The limit of the norm exits and is well defined (finite and non-vanishing)
	\begin{equation}
		\lim_{v_1 \rightarrow u_1} \lim_{v_2 \rightarrow u_2} \dots \, \, S_L (\{ u \} \rightarrow \{ v \}) =S_L (\u ) \ .
	\end{equation}
\end{itemize}
These are not the usual Smirnov's axioms for form factors because we are constructing the consistency condition from the ABA instead of from ZF operators, so the properties satisfied by both are different. In addition, because we are working in the weak coupling regime, we will not deal in this article with the constrains imposed by crossing transformations to this square.

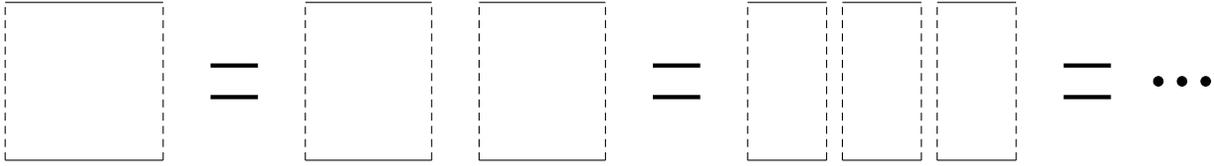
\begin{figure}[t]
\begin{center}
\begin{tikzpicture}[scale=2.1]
\draw [densely dashed] (0,0) --(0,1);
\draw (0, 0) -- (1,0);
\draw (0, 1) -- (1,1);
\draw [densely dashed] (1,0) --(1,1);
\draw [ultra thick] (1.3,0.4) -- (1.6,0.4);
\draw [ultra thick] (1.3,0.6) -- (1.6,0.6);
\draw [densely dashed] (1.9,0) --(1.9,1);
\draw (1.9, 0) -- (2.7,0);
\draw (1.9, 1) -- (2.7,1);
\draw [densely dashed] (2.7,0) --(2.7,1);
\draw [densely dashed] (3,0) --(3,1);
\draw (3, 0) -- (3.8,0);
\draw (3, 1) -- (3.8,1);
\draw [densely dashed] (3.8,0) --(3.8,1);
\draw [ultra thick] (4.1,0.4) -- (4.4,0.4);
\draw [ultra thick] (4.1,0.6) -- (4.4,0.6);
\draw [densely dashed] (4.7,0) --(4.7,1);
\draw (4.7, 0) -- (5.2,0);
\draw (4.7, 1) -- (5.2,1);
\draw [densely dashed] (5.2,0) --(5.2,1);
\draw [densely dashed] (5.3,0) --(5.3,1);
\draw (5.3, 0) -- (5.8,0);
\draw (5.3, 1) -- (5.8,1);
\draw [densely dashed] (5.8,0) --(5.8,1);
\draw [densely dashed] (5.9,0) --(5.9,1);
\draw (5.9, 0) -- (6.4,0);
\draw (5.9, 1) -- (6.4,1);
\draw [densely dashed] (6.4,0) --(6.4,1);
\draw [ultra thick] (6.7,0.4) -- (7,0.4);
\draw [ultra thick] (6.7,0.6) -- (7,0.6);
\draw [fill] (7.3,0.5) circle [radius=0.03];
\draw [fill] (7.45,0.5) circle [radius=0.03];
\draw [fill] (7.6,0.5) circle [radius=0.03];
\end{tikzpicture}
\caption{Pictorial representation of the cutting of an empty square.} \label{empty}
\end{center}
\end{figure}

The first case we should compute is the square with no excitations, which is fixed up to an unphysical normalization. The computation relies on the particular fact that such square can be cut as many times  as we want (as long as we are far from the discrete limit) involving no other structures. As we can see from figure~\ref{empty}, the consistency condition for it is given by
\begin{equation}
S_L (\emptyset )=S_{l_1} (\emptyset ) S_{l_2} (\emptyset )=S_{l'_1} (\emptyset ) S_{l'_2} (\emptyset ) S_{l'_3} (\emptyset )=\dots \ ,
\end{equation}
whose only physical solution is $S_L (\emptyset ) =1$. \footnote{Actually we can choose a more general solution $S_L (\emptyset )=\exp (\rho L)$ with $\rho$ being a constant. However, we are going to choose $\rho=0$ because there is no physical parameters with inverse length units we can use for that role.}

\begin{figure}[t]
\begin{center}
\begin{tikzpicture}[scale=2.1]
\draw [densely dashed] (0,0) --(0,1);
\draw (0, 0) -- (1,0);
\draw (0, 1) -- (1,1);
\draw [densely dashed] (1,0) --(1,1);
\draw [fill] (0.5,0) circle [radius=0.04];
\draw [fill] (0.5,1) circle [radius=0.04];
\draw [thick] (0.5,0) -- (0.5,1);
\draw [ultra thick] (1.3,0.4) -- (1.6,0.4);
\draw [ultra thick] (1.3,0.6) -- (1.6,0.6);
\draw [densely dashed] (1.9,0) --(1.9,1);
\draw (1.9, 0) -- (2.7,0);
\draw (1.9, 1) -- (2.7,1);
\draw [densely dashed] (2.7,0) --(2.7,1);
\draw [fill] (2.3,0) circle [radius=0.04];
\draw [fill] (2.3,1) circle [radius=0.04];
\draw [thick] (2.3,0) -- (2.3,1);
\draw [densely dashed] (2.9,0) --(2.9,1);
\draw (2.9, 0) -- (3.7,0);
\draw (2.9, 1) -- (3.7,1);
\draw [densely dashed] (3.7,0) --(3.7,1);

\draw [thin] (2.9,1.05) -- (2.9,1.15);
\draw [thin] (3.7,1.05) -- (3.7,1.15);
\draw [thin] (2.9,1.1) -- (3.7,1.1);
\node [above] at (3.3,1.1) {$\scriptstyle l_2$};

\draw [ultra thick] (4,0.5) -- (4.3,0.5);
\draw [ultra thick] (4.15,0.35) -- (4.15,0.65);
\draw [densely dashed] (4.6,0) --(4.6,1);
\draw (4.6, 0) -- (5.4,0);
\draw (4.6, 1) -- (5.4,1);
\draw [densely dashed] (5.4,0) --(5.4,1);

\draw [thin] (4.6,1.05) -- (4.6,1.15);
\draw [thin] (5.4,1.05) -- (5.4,1.15);
\draw [thin] (4.6,1.1) -- (5.4,1.1);
\node [above] at (5,1.1) {$\scriptstyle l_1$};

\draw [densely dashed] (5.6,0) --(5.6,1);
\draw (5.6, 0) -- (6.4,0);
\draw (5.6, 1) -- (6.4,1);
\draw [densely dashed] (6.4,0) --(6.4,1);
\draw [fill] (6,0) circle [radius=0.04];
\draw [fill] (6,1) circle [radius=0.04];
\draw [thick] (6,0) -- (6,1);
\end{tikzpicture}
\caption{Pictorial representation of the cutting of a square with one excitation.} \label{one}
\end{center}
\end{figure}
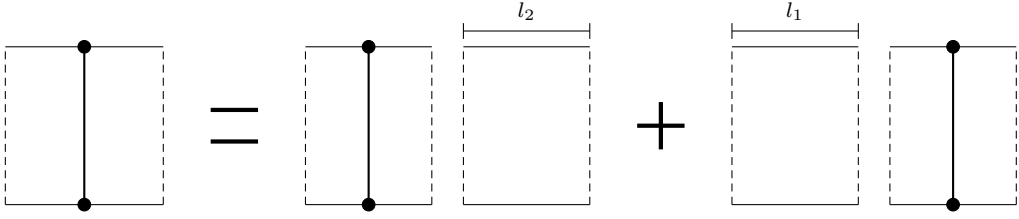

With the information about the empty square, we can move now to the computation of the square with one excitation. As we can see from its pictorial representation in figure~\ref{one}, equation~(\ref{bootstrap}) reflects the fact that the excitation can end in either of the two new squares we get after we cut the original one. Applied to this case, this equation reduces to
\begin{equation}
	S_{L} (u)=S_{l_1} (u)  S_{l_2} (\emptyset) a_{l_2} (u) d_{l_2} (u)+S_{l_1} (\emptyset)  S_{l_2} (u) a_{l_1} (u) d_{l_1} (u) \ .
\end{equation}
No other term can appear as we are not including mirror excitations. We can see that this equation simplifies by means of the rewriting $S_L (u)=\tilde{S}_L (u) a_L (u) d_L (u)$, reducing to
\begin{equation}
	\tilde{S}_L (u)=\tilde{S}_{l_1} (u) + \tilde{S}_{l_2} (u) \ .
\end{equation}
The only way to solve this equation is to have $\tilde{S}_L (u) =\sigma^{(1)}_1 (u) L$, where $\sigma^{(1)}_1$ is a general function not fixed by the constraint. Note that we cannot include a term independent of $L$ as the equation automatically fixes it to zero.

We can fix the structure of the $\sigma^{(1)}_1 (u)$ function if we compute the case in which the rapidity of the initial (lower-edge) excitation is different from the final (upper-edge) one and impose some analytic considerations. The consistency condition in this case reads
\begin{equation}
	S_{L} (u\rightarrow v)=S_{l_1} (u\rightarrow v)  S_{l_2} (\emptyset) a_{l_2} (v) d_{l_2} (u)+S_{l_1} (\emptyset)  S_{l_2} (u\rightarrow v) a_{l_1} (u) d_{l_1} (v) \ .
\end{equation}
To solve this equation we require the more general ansatz
\begin{equation}
	S_{L} (u\rightarrow v)= \tilde{S}_{L} (u\rightarrow v) \left[ a(u) d(v) - a(v) d(u) \right] \ ,
\end{equation}
which solves the consistency equation for any function $\tilde{S}_{L} (u\rightarrow v)$.

In order to get a non-vanishing norm we have to impose that $\tilde{S}_{L} (u\rightarrow v)$ has a pole of order one at $u=v$. This function can be further fixed through two different procedures. The first one is the direct comparison with the ABA matrix element $\bra{0} C(v) B(u) \ket{0}$ using (\ref{commCB}), which directly fixes $\tilde{S}_{L} (u\rightarrow v) =g(u,v)$. The second one is to use crossing symmetry. However, as we commented above, at weak coupling the crossing transformation is not accessible, so in order to extract the function from this symmetry we have to take the weak coupling limit after performing the transformation. As this involves dealing with the non-perturbative square, we have relegated the discussion on this topic to appendix~\ref{CrossingSquare}.

With the information from $S_{L} (u\rightarrow v)$ we can now extract the Gaudin norm associated to the one-excitation state
\begin{equation}
	\tilde{S}_L (u)= \frac{S_{L} (u)}{a(u) d(u)}=\lim_{v\rightarrow u} \tilde{S}_{L} (u\rightarrow v) \frac{d(v)}{a(u)} (e^{ip(v) L}- e^{ip(u) L} )= \text{const.} (-i p' L) \ ,
\end{equation}
where the momentum is defined as $ip(u)L=\log [a(u)/d(u)]$ and $p'=\frac{dp(u)}{du}$. We can see that the expression we have obtained agrees with the Gaudin norm (\ref{gaudin}) for one excitation.

\begin{figure}[t]
\begin{center}
\begin{tikzpicture}[scale=2.1]
\draw [densely dashed] (0,0) --(0,1);
\draw (0, 0) -- (1,0);
\draw (0, 1) -- (1,1);
\draw [densely dashed] (1,0) --(1,1);
\draw [fill] (0.33333,0) circle [radius=0.04];
\draw [fill] (0.33333,1) circle [radius=0.04];
\draw [thick] (0.33333,0) -- (0.333333,1);
\draw [fill=lightgray,lightgray] (0.66666,0) circle [radius=0.04];
\draw [fill=lightgray,lightgray] (0.66666,1) circle [radius=0.04];
\draw [thick,lightgray] (0.66666,0) -- (0.666666,1);
\draw [ultra thick] (1.3,0.4) -- (1.6,0.4);
\draw [ultra thick] (1.3,0.6) -- (1.6,0.6);
\draw [densely dashed] (1.9,0) --(1.9,1);
\draw (1.9, 0) -- (2.7,0);
\draw (1.9, 1) -- (2.7,1);
\draw [densely dashed] (2.7,0) --(2.7,1);
\draw [fill] (2.16666,0) circle [radius=0.04];
\draw [fill] (2.16666,1) circle [radius=0.04];
\draw [thick] (2.16666,0) -- (2.16666,1);
\draw [fill=lightgray,lightgray] (2.43333,0) circle [radius=0.04];
\draw [fill=lightgray,lightgray] (2.43333,1) circle [radius=0.04];
\draw [thick,lightgray] (2.43333,0) -- (2.43333,1);
\draw [densely dashed] (2.9,0) --(2.9,1);
\draw (2.9, 0) -- (3.7,0);
\draw (2.9, 1) -- (3.7,1);
\draw [densely dashed] (3.7,0) --(3.7,1);
\draw [ultra thick] (4,0.5) -- (4.3,0.5);
\draw [ultra thick] (4.15,0.35) -- (4.15,0.65);
\draw [densely dashed] (4.6,0) --(4.6,1);
\draw (4.6, 0) -- (5.4,0);
\draw (4.6, 1) -- (5.4,1);
\draw [densely dashed] (5.4,0) --(5.4,1);
\draw [densely dashed] (5.6,0) --(5.6,1);
\draw (5.6, 0) -- (6.4,0);
\draw (5.6, 1) -- (6.4,1);
\draw [densely dashed] (6.4,0) --(6.4,1);
\draw [fill] (5.86666,0) circle [radius=0.04];
\draw [fill] (5.86666,1) circle [radius=0.04];
\draw [thick] (5.86666,0) -- (5.86666,1);
\draw [fill=lightgray,lightgray] (6.13333,0) circle [radius=0.04];
\draw [fill=lightgray,lightgray] (6.13333,1) circle [radius=0.04];
\draw [thick,lightgray] (6.13333,0) -- (6.13333,1);

\draw [ultra thick] (1.35,0.5-1.5) -- (1.65,0.5-1.5);
\draw [ultra thick] (1.5,0.35-1.5) -- (1.5,0.65-1.5);
\draw [densely dashed] (1.9,0-1.5) --(1.9,1-1.5);
\draw (1.9, 0-1.5) -- (2.7,0-1.5);
\draw (1.9, 1-1.5) -- (2.7,1-1.5);
\draw [densely dashed] (2.7,0-1.5) --(2.7,1-1.5);
\draw [fill] (2.3,0-1.5) circle [radius=0.04];
\draw [fill] (2.3,1-1.5) circle [radius=0.04];
\draw [thick] (2.3,0-1.5) -- (2.3,1-1.5);
\draw [densely dashed] (2.9,0-1.5) --(2.9,1-1.5);
\draw (2.9, 0-1.5) -- (3.7,0-1.5);
\draw (2.9, 1-1.5) -- (3.7,1-1.5);
\draw [densely dashed] (3.7,0-1.5) --(3.7,1-1.5);
\draw [fill=lightgray,lightgray] (3.3,0-1.5) circle [radius=0.04];
\draw [fill=lightgray,lightgray] (3.3,1-1.5) circle [radius=0.04];
\draw [thick,lightgray] (3.3,0-1.5) -- (3.3,1-1.5);
\draw [ultra thick] (4,0.5-1.5) -- (4.3,0.5-1.5);
\draw [ultra thick] (4.15,0.35-1.5) -- (4.15,0.65-1.5);
\draw [densely dashed] (4.6,0-1.5) --(4.6,1-1.5);
\draw (4.6, 0-1.5) -- (5.4,0-1.5);
\draw (4.6, 1-1.5) -- (5.4,1-1.5);
\draw [densely dashed] (5.4,0-1.5) --(5.4,1-1.5);
\draw [fill=lightgray,lightgray] (5,0-1.5) circle [radius=0.04];
\draw [fill=lightgray,lightgray] (5,1-1.5) circle [radius=0.04];
\draw [thick,lightgray] (5,0-1.5) -- (5,1-1.5);
\draw [densely dashed] (5.6,0-1.5) --(5.6,1-1.5);
\draw (5.6, 0-1.5) -- (6.4,0-1.5);
\draw (5.6, 1-1.5) -- (6.4,1-1.5);
\draw [densely dashed] (6.4,0-1.5) --(6.4,1-1.5);
\draw [fill] (6,0-1.5) circle [radius=0.04];
\draw [fill] (6,1-1.5) circle [radius=0.04];
\draw [thick] (6,0-1.5) -- (6,1-1.5);
\end{tikzpicture}
\caption{Pictorial mnemonic representation of the cutting of a square with two excitations. Each color represents a different rapidity.} \label{two}
\end{center}
\end{figure}
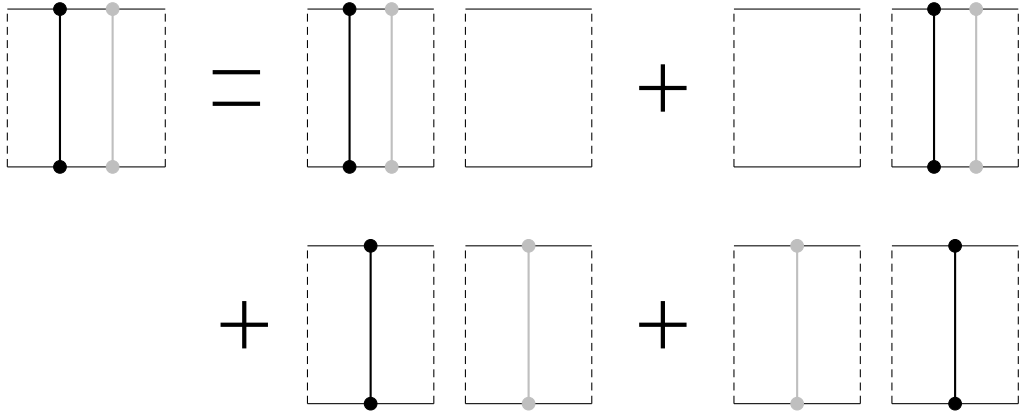

Moving now to the case of two excitations, equation~\ref{bootstrap} (pictorially represented in fig.~\ref{two}) now reads
\begin{align}
	S_{L}(u,v) &=S_{l_1}(u,v) S_{l_2} (\emptyset)  a_{l_2} (u) a_{l_2} (v) d_{l_2} (u) d_{l_2} (v)+S_{l_1} (\emptyset)  S_{l_2}(u,v) a_{l_1} (u) a_{l_1} (v) d_{l_1} (u) d_{l_1} (v) \notag \\ 
	&+S_{l_1} (u) S_{l_2} (v) f(u,v) f(v,u) a_{l_1} (v) d_{l_1} (v) a_{l_2} (u) d_{l_2} (u) \notag \\
	&+S_{l_1} (v) S_{l_2} (u) f(u,v) f(v,u) a_{l_1} (u) d_{l_1} (u) a_{l_2} (v) d_{l_2} (v)\ . \label{twomagnonansatz}
\end{align}
One might argue that there are two terms we have not taken into account, as we have not included terms involving squares whose excitations in the upper and lower sides have different rapidities. Although this is true in general, in this section we are trying to reconstruct the Gaudin norm and these terms vanish because $S_{L} (\{u\} \rightarrow \{v\neq u\})=0$ when the Bethe equations are imposed.

Following the same procedure we used for the case of one excitation,  we can factor out the momentum dependence, encoded in the functions $a$ and $d$, by defining $S_{L}(u,v)=\tilde{S}_{L}(u,v) a(u) a(v) d(u) d(v)$. However, in this case the ansatz for $\tilde{S}_{L}(u,v)$ has to be generalized to 
\begin{equation}
	\tilde{S}_{L}(u,v)= \sigma^{(2)}_2 (u,v) L^2 +\sigma^{(2)}_1 (u,v) L \ .
\end{equation}
Note again that adding an extra term with no dependence in $L$ is forbidden by the consistency condition. Substituting this ansatz into (\ref{twomagnonansatz}) fixes the first function to
\begin{equation}
	\sigma^{(2)}_2= f(u,v) f(v,u) \sigma^{(1)}_1 (u) \sigma^{(1)}_1 (v) \ ,
\end{equation}
but it does not give us any information about the second one. We can reconstruct the function $\sigma^{(2)}_1$ using the same trick as before, i.e., computing the square $S (\{u,v\} \rightarrow \{w,v\} )$ and taking the limit $w\rightarrow u$. Nevertheless, we are going to postpone this calculation to section~\ref{offshelltwoexcitations}, where we compute the off-shell square with two excitations, because such limit should be taken without imposing Bethe equations.

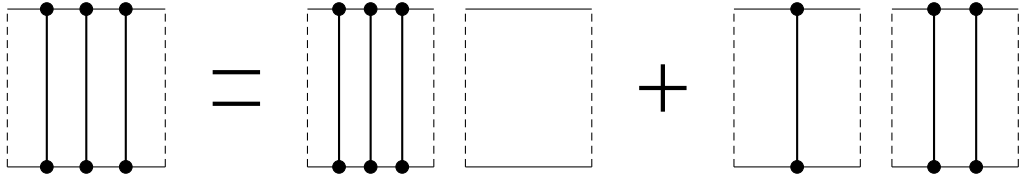
\begin{figure}[t]
\begin{center}
\begin{tikzpicture}[scale=2.1]
\draw [densely dashed] (0,0) --(0,1);
\draw (0, 0) -- (1,0);
\draw (0, 1) -- (1,1);
\draw [densely dashed] (1,0) --(1,1);
\draw [fill] (0.25,0) circle [radius=0.04];
\draw [fill] (0.25,1) circle [radius=0.04];
\draw [thick] (0.25,0) -- (0.25,1);
\draw [fill] (0.5,0) circle [radius=0.04];
\draw [fill] (0.5,1) circle [radius=0.04];
\draw [thick] (0.5,0) -- (0.5,1);
\draw [fill] (0.75,0) circle [radius=0.04];
\draw [fill] (0.75,1) circle [radius=0.04];
\draw [thick] (0.75,0) -- (0.75,1);
\draw [ultra thick] (1.3,0.4) -- (1.6,0.4);
\draw [ultra thick] (1.3,0.6) -- (1.6,0.6);
\draw [densely dashed] (1.9,0) --(1.9,1);
\draw (1.9, 0) -- (2.7,0);
\draw (1.9, 1) -- (2.7,1);
\draw [densely dashed] (2.7,0) --(2.7,1);
\draw [fill] (2.1,0) circle [radius=0.04];
\draw [fill] (2.1,1) circle [radius=0.04];
\draw [thick] (2.1,0) -- (2.1,1);
\draw [fill] (2.3,0) circle [radius=0.04];
\draw [fill] (2.3,1) circle [radius=0.04];
\draw [thick] (2.3,0) -- (2.3,1);
\draw [fill] (2.5,0) circle [radius=0.04];
\draw [fill] (2.5,1) circle [radius=0.04];
\draw [thick] (2.5,0) -- (2.5,1);
\draw [densely dashed] (2.9,0) --(2.9,1);
\draw (2.9, 0) -- (3.7,0);
\draw (2.9, 1) -- (3.7,1);
\draw [densely dashed] (3.7,0) --(3.7,1);
\draw [ultra thick] (4,0.5) -- (4.3,0.5);
\draw [ultra thick] (4.15,0.35) -- (4.15,0.65);
\draw [densely dashed] (4.6,0) --(4.6,1);
\draw (4.6, 0) -- (5.4,0);
\draw (4.6, 1) -- (5.4,1);
\draw [densely dashed] (5.4,0) --(5.4,1);
\draw [fill] (5,0) circle [radius=0.04];
\draw [fill] (5,1) circle [radius=0.04];
\draw [thick] (5,0) -- (5,1);
\draw [densely dashed] (5.6,0) --(5.6,1);
\draw (5.6, 0) -- (6.4,0);
\draw (5.6, 1) -- (6.4,1);
\draw [densely dashed] (6.4,0) --(6.4,1);
\draw [fill] (5.86666,0) circle [radius=0.04];
\draw [fill] (5.86666,1) circle [radius=0.04];
\draw [thick] (5.86666,0) -- (5.86666,1);
\draw [fill] (6.13333,0) circle [radius=0.04];
\draw [fill] (6.13333,1) circle [radius=0.04];
\draw [thick] (6.13333,0) -- (6.13333,1);
\end{tikzpicture}
\caption{Schematic representation of the cutting of a square with three excitations. The other structures are obtained by exchanging $l_1$ and $l_2$ and the labeling of rapidities, giving a total of 8 contributions.} \label{three}
\end{center}
\end{figure}

The last particular case we are going to dissect before presenting the general reconstruction is the one involving three excitations. In this case we can express equation~\ref{bootstrap}, schematically represented in fig.~\ref{three}, as
\begin{align}
	S_{L}(u,v,w) &=S_{l_1}(u,v,w)  S_{l_2} (\emptyset)  a_{l_2} (u) a_{l_2} (v) a_{l_2} (w) d_{l_2} (u) d_{l_2} (v) d_{l_2} (w) \notag \\
	&+S_{l_1} (\emptyset) S_{l_2}(u,v,w) a_{l_1} (u) a_{l_1} (v) a_{l_1} (w) d_{l_1} (u) d_{l_1} (v) d_{l_1} (w) \notag \\
	&+[S_{l_1} (u) S_{l_2} (v,w) f(u,v) f(u,w) f(v,u) f(w,u) a_{l_1} (v) d_{l_1} (v) a_{l_1} (w) d_{l_1} (w) a_{l_2} (u) d_{l_2} (u) \notag \\
	&+S_{l_1} (v,w) S_{l_2} (u) f(u,v) f(u,w) f(v,u) f(w,u) a_{l_1} (u) d_{l_1} (u) a_{l_2} (v) d_{l_2} (v) a_{l_2} (w) d_{l_2} (w)] \notag \\
	&+[u\leftrightarrow v] + [u \leftrightarrow w] \ ,
\end{align}
where the last two permutations are taken only over the terms in brackets.

To solve this consistency condition we need to substitute a polynomial of degree three in $L$ as ansatz. The reason behind it is that the equation involves the product of a square with one excitation (linear in $L$) and a square with two excitations (quadratic in $L$)
\begin{equation}
	S_{L}(u,v)=\left( \sigma^{(3)}_3 (u,v) L^3 + \sigma^{(3)}_2 (u,v) L^2 +\sigma^{(3)}_1 (u,v) L \right) a(u) a(v) a(w) d(u) d(v) d(w) \ .
\end{equation}
Similarly to the case of two excitations, the consistency condition fixes $\sigma^{(3)}_3$ and $\sigma^{(3)}_2$ but the function $\sigma^{(3)}_1$ remains unconstrained. In particular
\begin{align}
	\sigma^{(3)}_3 &= f(u,v)f(u,w) f(v,w) f(w,v) f(w,u) f(v,u) \sigma^{(1)}_1 (u) \sigma^{(1)}_1 (v) \sigma^{(1)}_1 (w) \ , \\
	\sigma^{(3)}_2 &= f(u,w) f(v,w) f(w,v) f(w,u)\sigma^{(2)}_1 (u,v)  \sigma^{(1)}_1 (w) +[u\leftrightarrow w] + [v \leftrightarrow w] \ .
\end{align}
This structure is very reminiscent of the Gaudin determinant rewritten using the Kirchhoff's matrix-tree theorem if we understand the $\sigma^{(N)}_1$ as functions associated to spanning trees of the complete graph with $N$ nodes. The relation between our results and the forest expansion will be clarified the following subsection.

\subsection{Reconstructing the Gaudin determinant: the general asymptotic square}

With the experience we have gathered from the previous particular cases, we are in a position to derive the full Gaudin determinant for the on-shell Bethe state with rapidities $\u=\{u_1 , \dots ,u_N\}$. The ansatz compatible with the consistency equation for the general case is
\begin{equation}
	S_{L} (\u)=a(\u) d(\u) \sum_{n=1}^{N} \sigma_n^{(N)} (\u) L^n \ .
\end{equation}
After substituting this ansatz into equation~\ref{bootstrap}, the resulting constrains can be solved by
\begin{equation}
	\sigma^{(N)}_i= f^{\neq} (\u , \u )  \sum_{\substack{\alpha_1 \cup \alpha_2 \cup \dots \cup \alpha_i =\u \\ \alpha_j \cap \alpha_k = \emptyset \\  \alpha_j \neq \emptyset}} \prod_{j=1}^i \frac{\sigma^{(|\alpha_j|)}_1 (\{\alpha_j \})}{f^{\neq} (\{\alpha_j\} , \{\alpha_j\})} \ .
\end{equation}
To clarify the formula, the explicit expressions for the two highest orders in length are
\begin{align}
	\sigma^{(N)}_N &=f^{\neq} (\u , \u ) \prod_{u_i \in \u} \sigma^{(1)}_1 (u_i) \ , \\
	\sigma^{(N)}_{N-1} &= \sigma^{(N)}_N  \sum_{\substack{u_i,u_j\in \u \\ u_i\neq u_j}}\frac{\sigma^{(2)}_1 (u_i,u_j) }{f(u_i,u_j) f(u_j,u_i) \sigma^{(1)}_1 (u_i) \sigma^{(1)}_1 (u_j)} \ .
\end{align}

As we said in the previous subsection, we can interpret this formula as the tree expansion of the Gaudin determinant. However, to be able to make this connection we still have to make another assumption. At this point, the $\sigma^{(N)}_i$ functions are the expressions associated to dividing the complete vertex-labelled graph with $N$ vertices into $i$ disconnected objects, but the individual $\sigma^{(|\alpha_i |)}_1$ still cannot be interpreted as a tree.

In order to proceed, we are going to assume that the $\sigma^{(n)}_1$ functions inherit certain factorization properties from the factorized scattering of the theory. In particular, we will assume that they can be written as the sum of the product of $n-1$ factors of $\sigma^{(2)}_1$ over non-repeated pairings of the $n$ arguments. This implies that we only have to fix the functions $\sigma^{(1)}_1$ and $\sigma^{(2)}_1$ to completely fix all squares, a task achievable just by using the ABA computations of the norm for one and two excitations as initial conditions. Even more, in the following subsection we will see that we can compute $\sigma^{(2)}_1$ just from the consistency condition, so we actually only need $\sigma^{(1)}_1$. A second consequence is that the sum over all possible pairings of the arguments can be understood as the sum over all possible spanning trees of the complete $n$-graph. The reason for that lies in the fact that the products involve $n-1$ non-repeated pairs, which is equivalent to the definition of a tree, can be mapped to the tree graph constructed with each vertex and each edge representing respectively a rapidity variable and a function $\frac{\sigma^{(2)}_1 (u,v)}{\sigma^{(1)}_1 (u) +\sigma^{(1)}_1 (v)}$ whose arguments are the vertices connected by the edge. Thus, allowing us to finally connect our formula with a forest expansion. We have chosen to divide $\sigma^{(2)}_1$ by the sum of the derivative of the momenta because it appears in the square with two magnons accompanied by a factor of the length, so it should contain a factor of the momentum and it should appear in a symmetric way. In the next section we will see that $\sigma^{(2)}_1 (u,v)\propto [p'(u) + p'(v)] \partial_{u} \log S(u,v)$. We can check from the expression~(\ref{treegaudin}) that our naive construction for the $\sigma^{(n)}_1$ just lacks a factor of the sum of the momenta of all excitations. \footnote{The sum over momenta appearing here can be made explicit in (\ref{treegaudin}) by separating the sum over directed forests into a sum over undirected forests and the contribution from the roots, being this last sum the sum over momenta.}

We still have to reconstruct the dependence of the two-point function with the spacetime. We discuss two different approaches to it in appendix~\ref{spacetimedependence}.

\section{Computing the off-shell square}

After being able to reconstruct the Gaudin norm using the consistency condition \ref{bootstrap}, in this section we are going to include all the possible asymptotic terms appearing on the equation to reconstruct the norm of off-shell Bethe states. We will detail the computations for the cases of two and three excitations. We will also compute the first non-Gaudin term appearing in the square with a general number of excitations. The case of the square with one excitation require no special treatment as no new terms appear in the consistency equation, hence it remains unchanged
\begin{equation}
	S_{L} (u\rightarrow v) =g(u,v) \left[ a(u) d(v) - a(v) d(u) \right] \ , \qquad S_{L} (u)= \text{const.} (-i p' L) e^{ipL} \ .
\end{equation}

\subsection{The off-shell square with two excitations}
\label{offshelltwoexcitations}

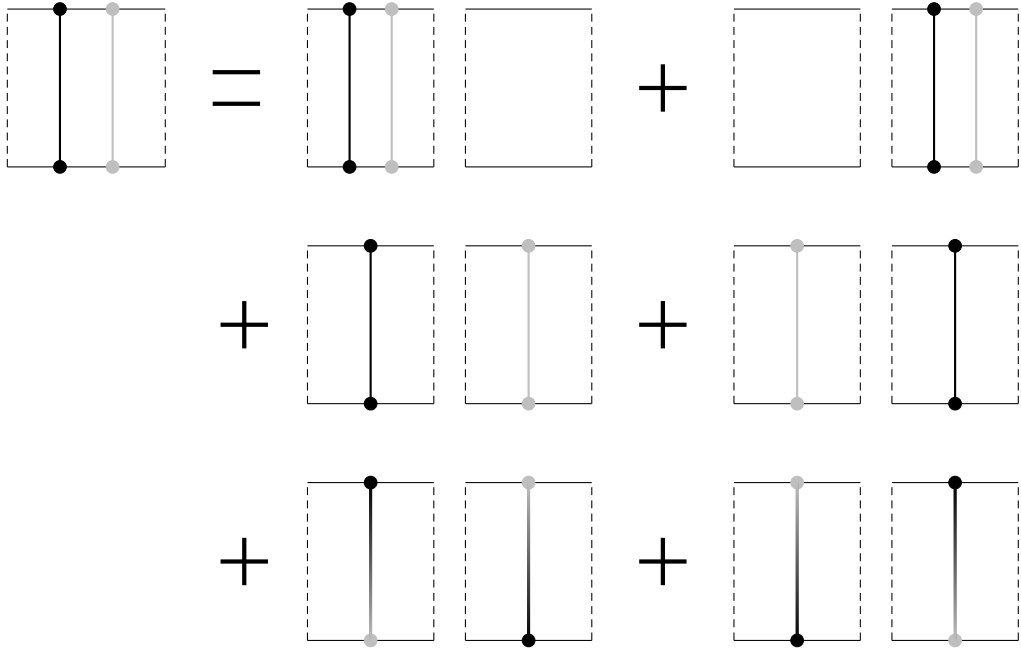
\begin{figure}[t]
\begin{center}
\begin{tikzpicture}[scale=2.1]
\draw [densely dashed] (0,0) --(0,1);
\draw (0, 0) -- (1,0);
\draw (0, 1) -- (1,1);
\draw [densely dashed] (1,0) --(1,1);
\draw [fill] (0.33333,0) circle [radius=0.04];
\draw [fill] (0.33333,1) circle [radius=0.04];
\draw [thick] (0.33333,0) -- (0.333333,1);
\draw [fill=lightgray,lightgray] (0.66666,0) circle [radius=0.04];
\draw [fill=lightgray,lightgray] (0.66666,1) circle [radius=0.04];
\draw [thick,lightgray] (0.66666,0) -- (0.666666,1);
\draw [ultra thick] (1.3,0.4) -- (1.6,0.4);
\draw [ultra thick] (1.3,0.6) -- (1.6,0.6);
\draw [densely dashed] (1.9,0) --(1.9,1);
\draw (1.9, 0) -- (2.7,0);
\draw (1.9, 1) -- (2.7,1);
\draw [densely dashed] (2.7,0) --(2.7,1);
\draw [fill] (2.16666,0) circle [radius=0.04];
\draw [fill] (2.16666,1) circle [radius=0.04];
\draw [thick] (2.16666,0) -- (2.16666,1);
\draw [fill=lightgray,lightgray] (2.43333,0) circle [radius=0.04];
\draw [fill=lightgray,lightgray] (2.43333,1) circle [radius=0.04];
\draw [thick,lightgray] (2.43333,0) -- (2.43333,1);
\draw [densely dashed] (2.9,0) --(2.9,1);
\draw (2.9, 0) -- (3.7,0);
\draw (2.9, 1) -- (3.7,1);
\draw [densely dashed] (3.7,0) --(3.7,1);
\draw [ultra thick] (4,0.5) -- (4.3,0.5);
\draw [ultra thick] (4.15,0.35) -- (4.15,0.65);
\draw [densely dashed] (4.6,0) --(4.6,1);
\draw (4.6, 0) -- (5.4,0);
\draw (4.6, 1) -- (5.4,1);
\draw [densely dashed] (5.4,0) --(5.4,1);
\draw [densely dashed] (5.6,0) --(5.6,1);
\draw (5.6, 0) -- (6.4,0);
\draw (5.6, 1) -- (6.4,1);
\draw [densely dashed] (6.4,0) --(6.4,1);
\draw [fill] (5.86666,0) circle [radius=0.04];
\draw [fill] (5.86666,1) circle [radius=0.04];
\draw [thick] (5.86666,0) -- (5.86666,1);
\draw [fill=lightgray,lightgray] (6.13333,0) circle [radius=0.04];
\draw [fill=lightgray,lightgray] (6.13333,1) circle [radius=0.04];
\draw [thick,lightgray] (6.13333,0) -- (6.13333,1);

\draw [ultra thick] (1.35,0.5-1.5) -- (1.65,0.5-1.5);
\draw [ultra thick] (1.5,0.35-1.5) -- (1.5,0.65-1.5);
\draw [densely dashed] (1.9,0-1.5) --(1.9,1-1.5);
\draw (1.9, 0-1.5) -- (2.7,0-1.5);
\draw (1.9, 1-1.5) -- (2.7,1-1.5);
\draw [densely dashed] (2.7,0-1.5) --(2.7,1-1.5);
\draw [fill] (2.3,0-1.5) circle [radius=0.04];
\draw [fill] (2.3,1-1.5) circle [radius=0.04];
\draw [thick] (2.3,0-1.5) -- (2.3,1-1.5);
\draw [densely dashed] (2.9,0-1.5) --(2.9,1-1.5);
\draw (2.9, 0-1.5) -- (3.7,0-1.5);
\draw (2.9, 1-1.5) -- (3.7,1-1.5);
\draw [densely dashed] (3.7,0-1.5) --(3.7,1-1.5);
\draw [fill=lightgray,lightgray] (3.3,0-1.5) circle [radius=0.04];
\draw [fill=lightgray,lightgray] (3.3,1-1.5) circle [radius=0.04];
\draw [thick,lightgray] (3.3,0-1.5) -- (3.3,1-1.5);
\draw [ultra thick] (4,0.5-1.5) -- (4.3,0.5-1.5);
\draw [ultra thick] (4.15,0.35-1.5) -- (4.15,0.65-1.5);
\draw [densely dashed] (4.6,0-1.5) --(4.6,1-1.5);
\draw (4.6, 0-1.5) -- (5.4,0-1.5);
\draw (4.6, 1-1.5) -- (5.4,1-1.5);
\draw [densely dashed] (5.4,0-1.5) --(5.4,1-1.5);
\draw [fill=lightgray,lightgray] (5,0-1.5) circle [radius=0.04];
\draw [fill=lightgray,lightgray] (5,1-1.5) circle [radius=0.04];
\draw [thick,lightgray] (5,0-1.5) -- (5,1-1.5);
\draw [densely dashed] (5.6,0-1.5) --(5.6,1-1.5);
\draw (5.6, 0-1.5) -- (6.4,0-1.5);
\draw (5.6, 1-1.5) -- (6.4,1-1.5);
\draw [densely dashed] (6.4,0-1.5) --(6.4,1-1.5);
\draw [fill] (6,0-1.5) circle [radius=0.04];
\draw [fill] (6,1-1.5) circle [radius=0.04];
\draw [thick] (6,0-1.5) -- (6,1-1.5);

\draw [ultra thick] (1.35,0.5-3) -- (1.65,0.5-3);
\draw [ultra thick] (1.5,0.35-3) -- (1.5,0.65-3);
\draw [densely dashed] (1.9,0-3) --(1.9,1-3);
\draw (1.9, 0-3) -- (2.7,0-3);
\draw (1.9, 1-3) -- (2.7,1-3);
\draw [densely dashed] (2.7,0-3) --(2.7,1-3);
\draw [fill=lightgray,lightgray] (2.3,0-3) circle [radius=0.04];
\draw [fill] (2.3,1-3) circle [radius=0.04];
\draw [test={0.6pt}{lightgray}{black}, thick] (2.3,0-3) -- (2.3,1-3);
\draw [densely dashed] (2.9,0-3) --(2.9,1-3);
\draw (2.9, 0-3) -- (3.7,0-3);
\draw (2.9, 1-3) -- (3.7,1-3);
\draw [densely dashed] (3.7,0-3) --(3.7,1-3);
\draw [fill] (3.3,0-3) circle [radius=0.04];
\draw [fill=lightgray,lightgray] (3.3,1-3) circle [radius=0.04];
\draw [test={0.6pt}{black}{lightgray}, thick] (3.3,0-3) -- (3.3,1-3);
\draw [ultra thick] (4,0.5-3) -- (4.3,0.5-3);
\draw [ultra thick] (4.15,0.35-3) -- (4.15,0.65-3);
\draw [densely dashed] (4.6,0-3) --(4.6,1-3);
\draw (4.6, 0-3) -- (5.4,0-3);
\draw (4.6, 1-3) -- (5.4,1-3);
\draw [densely dashed] (5.4,0-3) --(5.4,1-3);
\draw [fill] (5,0-3) circle [radius=0.04];
\draw [fill=lightgray,lightgray] (5,1-3) circle [radius=0.04];
\draw [test={0.6pt}{black}{lightgray}, thick] (5,0-3) -- (5,1-3);
\draw [densely dashed] (5.6,0-3) --(5.6,1-3);
\draw (5.6, 0-3) -- (6.4,0-3);
\draw (5.6, 1-3) -- (6.4,1-3);
\draw [densely dashed] (6.4,0-3) --(6.4,1-3);
\draw [fill=lightgray,lightgray] (6,0-3) circle [radius=0.04];
\draw [fill] (6,1-3) circle [radius=0.04];
\draw [test={0.6pt}{lightgray}{black}, thick] (6,0-3) -- (6,1-3);
\end{tikzpicture}
\caption{Pictorial representation of the cutting process applied to a square with two excitations. Each color represents a different rapidity.} \label{twooff}
\end{center}
\end{figure}

The consistency condition for the case of two excitations, pictorially represented in fig.~\ref{twooff}, needs to include the two extra terms involving ``no norm-like'' one-excitation squares we neglected in the previous section, being its complete expression
\begin{align}
	S_L (u,v) &= S_{l_1} (u,v) a_{l_2} (u) a_{l_2} (v) d_{l_2} (u) d_{l_2} (v)+ S_{l_2} (u,v) a_{l_1} (u) a_{l_1} (v) d_{l_1} (u) d_{l_1} (v) + \notag \\
	&+S_{l_1} (u) S_{l_2} (v) f(u,v) f(v,u) a_{l_1} (v) d_{l_1} (v) a_{l_2} (u) d_{l_2} (u) \notag \\
	&+S_{l_1} (v) S_{l_2} (u) f(u,v) f(v,u) a_{l_1} (u) d_{l_1} (u) a_{l_2} (v) d_{l_2} (v) \notag \\
	&+S_{l_1} (v\rightarrow u) S_{l_2} (u\rightarrow v) f(v,u)^2 a_{L} (u) d_{L} (v) \notag \\
	&+S_{l_1} (u\rightarrow v) S_{l_2} (v\rightarrow u) f(u,v)^2 a_{L} (v) d_{L} (u)  \ .
\end{align}
As the Gaudin norm solves this equation except for the last two terms, we are going to look for a solution of the form
\begin{equation}
	S_L (u,v) = \tilde{S}^{(\text{Gaudin})}_L (u,v) a(u) a(v) d(u) d(v)+\text{extra terms} \ .
\end{equation}
It is important to remark that these new terms do not even have to have the same momentum structure as the on-shell terms. Indeed
\begin{multline}
	S_{l_1} (u\rightarrow v) S_{l_2} (v\rightarrow u) a_{L} (u) d_{L} (v) = -g(u,v)^2 \Big[ a_{L}(v) d_{L}(u) +a_{L}(v) d_{L}(u) \\
	- a_{l_1}(u) a_{l_2}(v) d_{l_2}(u)  d_{l_1}(v) - a_{l_2}(u) a_{l_1}(v) d_{l_1}(u) d_{l_2}(v) \Big] a_{L} (v) d_{L} (u) \ ,
\end{multline}
cannot be compensated by such momentum structure. We can upgrade our ansatz to the following one
\begin{align}
	S_L (u,v) &= \alpha_{u,v} [a_L (u) d_L (v) ]^ 2 + \alpha_{v,u} [a_L (v) d_L (u) ]^ 2 \notag \\
	&+ \left[ \sigma^{(2)}_2 L^2 +\sigma^{(2)}_1 (u,v) L +\sigma^{(2)}_0 \right] a_L (u) a_L (v) d_L (u) d_L (v) \ ,
\end{align}
where the subindexes of the function $\alpha$ indicates the structure of the momentum terms accompanying them. Thus, the consistency condition can be divided into seven different equations according to the different momentum factor accompanying each term \footnote{Note that the first six equations reduce to only two and the condition that $\alpha_{u,v}$ and $\alpha_{v,u}$ are independent of the length.}
\begin{align*}
	\alpha_{u,v}^{(L)} &= f(v,u)^2 g(u,v)^2 \ , & \alpha_{u,v}^{(l_1)} &= f(u,v)^2 g(u,v)^2  \ , & \alpha_{u,v}^{(l_2)} &= f(u,v)^2 g(u,v)^2  \ , \notag \\
	\alpha_{v,u}^{(L)} &= f(u,v)^2 g(v,u)^2 \ , & \alpha_{v,u}^{(l_1)} &= f(v,u)^2 g(v,u)^2  \ , & \alpha_{v,u}^{(l_2)} &= f(v,u)^2 g(v,u)^2  \ ,
\end{align*}
\begin{align}
	&\sigma^{(2)}_2 L^2 +\sigma^{(2)}_1 L +\sigma^{(2)}_0=\sigma^{(2)}_2 l_1^2 +\sigma^{(2)}_0+\sigma^{(2)}_2 l_2^2 \sigma^{(2)}_1 l_1+\sigma^{(2)}_1 l_2+\sigma^{(2)}_0  \\
	&+2l_1 l_2 \, \sigma_1^{(1)} (u) \sigma_1^{(1)} (v) f(u,v) f(v,u)+ g(u,v)^2 [f(u,v)^2 + f(v,u)^2] \ .\notag 
\end{align}
In contrast with the Gaudin square, here the term $\sigma_0$ is not only allowed, but needed and completely fixed. These equations are solved by
\begin{align}
	\alpha_{u,v} &=f(v,u)^2 g(u,v)^2 \ , & \sigma^{(2)}_2=f(u,v) f(v,u) \sigma_1^{(1)} (u) \sigma_1^{(1)} (v) \ , \notag \\
	 \alpha_{v,u} &=f(u,v)^2 g(u,v)^2 \ , & \sigma^{(2)}_0= -g(u,v)^2 [f(u,v)^2 +f(v,u)^2] \ . \label{sigmatwomagnonoff}
\end{align}
Again the function $\sigma^{(2)}_1 (u,v)$ is not fixed by this procedure.

Working with off-shell states we can constrain the function $\sigma^{(2)}_1 (u,v)$ in a similar way as we constrained the function $\sigma^{(1)}_1 (u)$ in the previous section, that is, by computing the square $S_L (\{ u,v \} \rightarrow \{ w,v \} )$ and taking the limit $w\rightarrow v$. Although we only have to compute the terms proportional to $p' (v)$ to extract this function, we will compute the full square as it will appear in the consistency equation for a square with three excitations. The consistency condition for this square is
\begin{align}
	&S_L (\{ u,v \} \rightarrow \{ w,v \} ) = S_{l_1} (\{ u,v \} \rightarrow \{ w,v \}) a_{l_2} (w) a_{l_2} (v) d_{l_2} (u) d_{l_2} (v) +  S_{l_2} (\{ u,v \} \rightarrow \{ w,v \} ) \notag \\
	& \cdot a_{l_1} (u) a_{l_1} (v) d_{l_1} (w) d_{l_1} (v)+S_{l_1} (u\rightarrow w) S_{l_2} (v) a_{l_2} (w) a_{l_1} (v) d_{l_2} (u) d_{l_1} (v) f(u,v) f(v,w)  \notag \\
	&+S_{l_1} (v) S_{l_2} (u\rightarrow w) a_{l_1} (u) a_{l_2} (v) d_{l_1} (w) d_{l_2} (v) f(w,v) f(v,u) \notag \\
	&+S_{l_1} (u\rightarrow v) S_{l_2} (v\rightarrow w) a_{l_1} (v) a_{l_2} (v) d_{l_2} (u) d_{l_1} (w) f(u,v) f(w,v) \notag \\
	&+S_{l_1} (v\rightarrow w) S_{l_2} (u\rightarrow v) a_{l_1} (u) a_{l_2} (w) d_{l_1} (v) d_{l_2} (v) f(v,u) f(v,w) \ .
\end{align}
Here we have to choose the following ansatz
\begin{align}
	S_L (\{ u,v \} \rightarrow \{ w,v \} ) &=\beta_u (v) a (u) a (v) d(v) d(w) +\beta_w (v) a (w) a(v) d(v) d(u) \notag \\
	&+ \alpha_a a (u) a(w) [d(v)]^2 +\alpha_d [a (v)]^2 d(u) d(w) \ .
\end{align}
Substituted into the consistency condition it gives us nine equations, which can be reduced to only five equations and the condition that $\alpha_a$ and $\alpha_d$ are independent of the length. These equations are solved by
\begin{align}
	\beta_u (v) &=g(u , w) f(w,v ) f(v, u) \sigma^{(1)}_1 (v) L +g(u,v) g(v,w)- \left[ g(u,v) g(v,w) \right]^2 +k(u,v,w) \ , \notag \\
	\beta_w (v) &=g(w , u) f(u, v) f(v,w) \sigma^{(1)}_1 (v) L +g(u,v) g(v,w)- \left[ g(u,v) g(v,w) \right]^2 -k(u,v,w) \ , \notag \\
	\alpha_a &= -g(u,v) g(v,w) f(v,u) f(v,w) \ , \notag \\
	\alpha_d &= -g(u,v) g(v,w) f(u,v) f(w,v) \  \label{beta2magnonoff}.
\end{align}
Note that there is a length-independent contribution to the functions $\beta$, which we label here as $k(u,v,w)$, that cannot be fixed from the consistency condition because these relations only involves the combination $\beta_{u,l_1} (v) + \beta_{w,l_2} (v)$. We may think that going one step further and computing $S_L(\{u_1 ,u_2 \} \rightarrow \{v_1 , v_2\} )$ would give us an equation that fixes this unknown function, but only an equivalent combination appears and the function cannot be fixed.

Part of $\sigma^{(2)}_1$ is extracted from this result by performing the limit $w\rightarrow u$ of the functions $\beta$ with their corresponding momentum factor, which needs some care as the $g(u,w)$ factors will transform into derivatives. Focusing only in these factors, divided by $\sigma^{(1)}_1 (v) L \, a(v) d(v)$, we have
\begin{multline}
	\lim_{w\rightarrow u}g(u,w)  \Big[ f(w,v) f(v,u) a (u) d(w) - f(u,v) f(v,w) a(w) d(u) \Big]= \\
	=\left( \frac{\partial f(u,v)}{\partial u} d(u) + f(u,v) \frac{\partial d(u)}{\partial u} \right) f(v,u) a(u) - \left( \frac{\partial f(v,u)}{\partial u} a(u) + f(v,u) \frac{\partial a(u)}{\partial u} \right) f(u,v) d(u) \ ,
\end{multline}
here the first term of each parenthesis together will generate half of $\sigma^{(2)}_1$ while the second ones generate $\sigma^{(2)}_2$. We can rewrite the two terms contributing to $\sigma^{(2)}_1$ as a single term
\begin{equation}
	\sigma^{(2)}_1 \propto [f(v,u)]^2 \frac{\partial \left( \frac{f(u,v)}{f(v,u)}\right)}{\partial u} \ ,
\end{equation}
which highlights that it is just the derivative of the S-matrix, as expected from the known expression of the Gaudin norm. The other half of $\sigma^{(2)}_1$  comes from the same limit applied to the $\alpha$ terms, giving the same result but dressed with $\sigma^{(1)}_1 (u)$ instead.

To end the subsection, we want to show that the non-Gaudin terms contain the two-excitation Bethe equations. This can be made explicit if we perform the rewriting
\begin{align}
	&\left[ a(u) d(v) f(v,u) g(u,v) \right] ^2 +\left[ a(v) d(u) f(u,v) g(v,u) \right] ^2 +a(u) a (v) d(u) d(v)  g(u,v) g(v,u) \notag \\
	&\cdot [f(u,v)^2 +f(v,u)^2]= a(u) a (v) d(u) d(v) [g(u,v)]^2 f(u,v) f(v,u) \left[ \frac{a(u)}{d(u)} \, \frac{d(v)}{a(v)} \, \frac{f(v,u)}{f(u,v)} \right. \notag \\
	&\left.- \frac{f(u,v)}{f(v,u)} -\frac{f(v,u)}{f(u,v)} +\frac{d(u)}{a(u)} \, \frac{a(v)}{d(v)} \, \frac{f(u,v)}{f(v,u)} \right] \ .
\end{align}
We can see that, apart from the vanishing inherited from the one-magnon square, these terms also vanish when the two-mangon Bethe equations (\ref{BAE})
\begin{equation}
	\frac{a(u)}{d(u)}=\frac{d(v)}{a(v)}=\frac{f(u,v)}{f(v,u)} \ ,
\end{equation}
are fulfilled, proving that the reconstruction is consistent with the Bethe Ansatz.

\subsection{The off-shell square with three excitations}

The computation of the square for the case of three excitations follows the same pattern as the previous case, but the large number of different combinations of momenta obscures the consistency condition. However, our experience from isolating the different momentum contributions in the two-excitation case had shown us that most of the equations we get are redundant. If we substitute the ansatz
\begin{align}
	&S_L (u,v,w) = \alpha_{u,v} [a_l (u) d_L (v) ]^ 2 a_L (w) d_L (w) + \alpha_{v,u} [a_L (v) d_L (v) ]^ 2 a_L (w) d_L (w)+ (w\leftrightarrow u)  \notag \\
	&+ (w\leftrightarrow v)+ \left[ \sigma^{(3)}_3 L^3 + \sigma^{(3)}_2 L^2 +\sigma^{(3)}_1 L +\sigma^{(3)}_0 \right] a_L (u) a_L (v) a_L (w) d_L (u) d_L (v) d_L (w) \ ,
\end{align}
into the consistency condition, only three sets of equations are relevant while the others can be extracted from symmetry under exchange of rapidities or independence of the length.

One set of equations we have to study is the set associated to the term $a(\u) d(\u)$, which codifies most of the length-dependent structure of the square. It is also important to study these equations because of the presence in the consistency condition of terms with the structure $\sigma^{(1)}_1 \sigma^{(2)}_0 L$, which were absent in the consistency condition for ``Gaudin squares'' and would make the equation inconsistent. Indeed, since the equation for $\sigma^{(3)}_1$ takes the form
\begin{equation}
	\sigma^{(3)}_1 L=\sigma^{(3)}_1 l_1 + \sigma^{(3)}_1 l_2 +h(u,v,w) L
\end{equation}
a solution does not exist unless the function $h(u,v,w)$ vanishes. \footnote{This is true for any power of $L$, as the equation $\sigma^{(n)}_j L^j=\sigma^{(n)}_j l_1^j + \sigma^{(n)}_j l_2^j +\sum_{l=1}^{j-1} \binom{j}{i} \alpha\left( \u \right) \, l_1^{j-l} l_2^l +h(u,v,w) L^j$ cannot be solved unless $h(u,v,w)$ vanishes.} The extra contribution in this equation is given by
\begin{align}
	&f(u,w) f(v,u) [f(v,w)]^2 g(v,w) \beta_v (u) +f(u,v) f(w,u) [f(w,v)]^2 g(w,v) \beta_w (u) \notag \\
	&+f(u,v) f(u,w) f(v,u) f(w,u) \sigma^{(1)}_1 (u) \sigma^{(2)}_0 (v,w) + (u\leftrightarrow v) + (u\leftrightarrow w) \ ,
\end{align}
which we can be shown to vanish when the explicit expression for $\beta$ and $\sigma$ from equations (\ref{sigmatwomagnonoff}) and (\ref{beta2magnonoff}) are substituted, as requires by consistency. Thus the equation we get just shows us that the function $\sigma_1^{(3)}$ is not fixed by the consistency condition.

We should remark two additional points regarding this set of equations. First, the orders $L^3$ and $L^2$ receive no contributions from the no norm-like squares, hence $\sigma^{(3)}_3$ and $\sigma^{(3)}_2$ are not modified with respect to the Gaudin case. And second, as the equation that fixes $\sigma^{(3)}_0$ involves the  function $k(u,v,w)$, appearing from the length-independent structure of the functions $\beta$, we need a larger set of inputs to completely fix the general square than for the Gaudin case.

The other equations we need to solve are the ones obtained from terms involving the momentum factors $a(\u) d(\u) \frac{a(u)}{d(u)} \frac{d(v)}{a(v)}$ and $a(\u) d(\u) \frac{d(u)}{a(u)} \frac{a(v)}{d(v)}$. Any other combination of rapidities is obtained by means of the symmetry inherited from the exchange symmetry of the $B$ and $C$ operators (\ref{commBB}). There equations are
\begin{align*}
	\alpha_{u,v} &= g(u,v) [f(v,u)]^2 f(w,u) f(v,w) \beta_u (w) \ , & \alpha_{v,u} &=g(u,v) [f(u,v)]^2 f(u,w) f(w,v) \beta_v (w) \ .
\end{align*}
Substituting the expression for the functions $\beta$ from equation (\ref{beta2magnonoff}) we get
\begin{align}
	\alpha_{u,v} &= [g(u,v) f(v,u) f(w,u) f(v,w)]^2 \sigma^{(1)}_1 (w) L - g(u,w) g(w,v) g(v,u) [f(v,u)]^2 f(w,u) f(v,w) \notag \\
	&+[g(u,w) g(w,v)]^2 g(v,u) [f(v,u)]^2 f(w,u) f(v,w)+ g(u,v) [f(v,u)]^2 f(w,u) f(v,w) k(u,w,v) \ , \notag \\
	\alpha_{v,u} &=-[g(u,v) f(u,v) f(u,w) f(w,v)]^2 \sigma^{(1)}_1 (w) +g(v,w) g(w,u) g(u,v) [f(u,v)]^2 f(u,w) f(w,v) \notag \\
	&-[g(v,w) g(w,u)]^2 g(u,v) [f(u,v)]^2 f(u,w) f(w,v)-g(u,v) [f(u,v)]^2 f(u,w) f(w,v) k(u,w,v) \ ,
\end{align}
which matches the norm computed using the commutation relations for the $B$ and $C$ operators at order $L$.

\subsection{The general off-shell norm-like square}

As we have seen in the previous case, the difficulty in computing the off-shell square and the normalization of off-shell Bethe states increases heavily with respect to the on-shell case. We are not able to give a full general expression for the off-shell square in this article, but we can at least give the exact expression for the first no norm-like contribution, which appears in the third highest order term in length. In order to do so it is more convenient to rewrite equation~\ref{bootstrap} as
\begin{multline}
	S_L (\{ u \} )= \sum_{\alpha \cup \beta \cup \gamma \cup \delta =\{u\}} w \Big( \{\alpha \cup \gamma\}, \{\beta \cup \delta \} , \{\alpha \cup \delta\} , \{ \beta \cup \gamma \} \Big) \\
	\cdot S_{l_1} \Big( \{\alpha \cup \gamma\} \rightarrow \{ \alpha \cup \delta\} \Big) S_{l_2} \Big( \{\beta \cup \delta\} \rightarrow \{ \beta \cup \gamma\} \Big) \ .
\end{multline}

A factor of $L$ appears for each pair of upper- and lower-edge excitations with the same rapidity, hence in the large length limit we have $S_{l_1} \Big( \{\alpha \cup \gamma\} \rightarrow \{ \alpha \cup \delta\} \Big) \propto l_1^{|\{\alpha\}|}$ and similarly for $S_{l_2}$ and $\{\beta\}$. The terms in the consistency condition involving the highest powers of the length come from the terms of the sum with coinciding outgoing and incoming rapidities, i.e. from the terms with $\delta=\gamma=\emptyset$. As a consequence, to get the highest and second highest contributions in length we only have to take into account norm-like squares, and thus these contributions coincide with the corresponding terms appearing in Gaudin norm. The third and fourth orders get extra contributions apart from the ones coming from the Gaudin norm. In particular, these extra contributions come from the terms in the consistency condition in which both squares have one excitation different in the upper edge with respect to the lower edge ($|\delta|=|\gamma|=1$, so its highest power has two powers of $L$ less than the highest power of the Gaudin contribution, one from each square).

Nevertheless, since the consistency condition involves the expression for the leading order in length of the square with one excitation different in the outgoing state with respect to the ingoing one, $S_{L} (\{u_i \cup \alpha\}\rightarrow \{u_j \cup \alpha\})$, we have to find its expression first. The correct ansatz in this case is
\begin{equation}
	S_L \left( \{u\} \rightarrow \left\{v \cup \bar{u}_1 \right\} \right)= a(\{u\} ) d(\{u\} ) \left[ \beta_{1,v} (\{ \bar{u}_1 \}) \frac{d(v)}{d(u_1)} + \beta_{v,1} (\{ \bar{u}_1 \}) \frac{a(v)}{a(u_1)} +\mathcal{O} (L^{N-1})  \right] \ ,
\end{equation}
where $\{ \bar{u}_1 \}$ stands for the set $\{u\}$ without the element $u_1$. We have already computed this square for two excitations and we have seen that it is proportional to the Gaudin norm of the excitation that is present in both physical sides. Inspired by this, we are going to propose that the functions $\beta_{1,v} (\{ \bar{u}_1 \})$ and $\beta_{v,1} (\bar{u}_1)$ are proportional to $\tilde{S}^{(\text{Gaudin})} (\{ \bar{u}_1 \})$ and check it by induction.

The leading order in length of the equation in the consistency condition associated to the ratio $\frac{a(v)}{a(u_1)}$ is
\begin{multline}
	\beta_{v,1}^{(L)} (\bar{u}_1) =\beta_{v,1}^{(l_1)} (\bar{u}_1) +g(v,u_1)  f(u_1 ,\bar{u}_1) f(\bar{u}_1 , v) \tilde{S}_{l_2} (\bar{u}_1) \\
	+\sum_{\substack{\alpha \cup \beta =\bar{u}_1 \\ \alpha \neq \emptyset}} f(u_1 , \beta) f(\beta , v) f(\alpha ,\beta ) f(\beta , \alpha) \beta^{l_1}_{v,1} (\alpha) \tilde{S}_{l_2} (\beta) \ .
\end{multline}
Thus, we can check that assuming the following structure for $|\alpha|<|\bar{u}_1|$
\begin{align}
	\beta^{(L)}_{1,v} (\alpha)&=f(\{ \alpha \}, u_1) f(v,\{ \alpha \}) g(u_1,v) \tilde{S}^{(Gaudin)}_L (\{ \alpha \}) \ , \notag \\
	\beta^{(L)}_{v,1} (\alpha)&=-f(\{ \alpha \}, v) f(u_1,\{ \alpha \}) g(u_1,v) \tilde{S}^{(Gaudin)}_L (\{ \alpha \}) \ .
\end{align}
gives an equivalent structure for $\beta^{(L)}_{1,v} (\bar{u}_1)$ and $\beta^{(L)}_{v,1} (\bar{u}_1)$, as the consistency condition can be recast into the consistency condition of the Gaudin square.

Now we can compute the first non-Gaudin contribution to third order in length of the off-shell square. The correction is generated from the terms $S_{l_1} (\{u_i \cup \alpha\}\rightarrow \{u_j \cup \alpha\}\} \cdot S_{l_2} (\{u_j \cup \beta\}\rightarrow \{u_i \cup \beta\}\}$ and the one obtained by exchanging $l_1$ and $l_2$, which can be divided into the following three momentum contributions
\begin{align}
&a(u_i) d(u_j) [f(u_i,u_j)]^2 f(u_i ,\bar{u}_{ij}) f(\bar{u}_{ij} , u_j) f(\alpha , \beta) \left[ a(u_j) d(u_i) \beta^{(l_1)}_{j,i} (\alpha) \beta^{(l_2)}_{j,i} (\beta) \right. \notag \\
&+\left. a(u_i) d(u_j) \beta^{(l_1)}_{i,j} (\alpha) \beta^{(l_2)}_{i,j} (\beta) \right] +a(u_j) d(u_i) [f(u_j,u_i)]^2 f(u_j ,\bar{u}_{ij}) f(\bar{u}_{ij} , u_i) \notag \\
&\cdot \left[ a(u_i) d(u_j) \beta^{(l_1)}_{i,j} (\beta) \beta^{(l_2)}_{i,j} (\alpha) +a(u_j) d(u_j)  \beta^{(l_1)}_{j,i} (\beta) \beta^{(l_2)}_{j,i} (\alpha)\right] \ .
\end{align}
For each different contribution we can reconstruct the consistency condition for $\tilde{S}^{(\text{Gaudin})}$, allowing us to write
\begin{align}
	S_L(\{u\}) &= S^{(\text{Gaudin})}_L(\{u\}) + \sum_{i\neq j} \left\{ \vphantom{\frac{a(u_i) d(u_j)}{a(u_j) d(u_i)} } \left[ g(u_i,u_j) f(u_j , u_i) f(\bar{u}_{ij} , u_j) f(u_i,\bar{u}_{ij}) \right] ^2 \right. \notag \\
	&+\left[ g(u_i,u_j) f(u_i , u_j) f(\bar{u}_{ij} , u_i) f(u_j,\bar{u}_{ij}) \right] ^2 + \left[ \frac{a(u_i) d(u_j)}{a(u_j) d(u_i)} \left[ f(u_i , u_j) g(u_i , u_j) \right]^2  \right. \notag \\
	&\cdot \left. \left. f(u_i , \bar{u}_{ij} ) f(u_j , \bar{u}_{ij} ) f(\bar{u}_{ij}, u_i ) f(\bar{u}_{ij}, u_j ) + i\leftrightarrow j \vphantom{\frac{a(u_i) d(u_j)}{a(u_j) d(u_i)} }\right]  \right\} a(\{u\}) d(\{u\}) \tilde{S}^{(\text{Gaudin})} (\bar{u}_{ij} ) \notag \\
	&+\mathcal{O} \left( L^{|\{u\}|-4} \right) \ .
\end{align}

\section{Conclusions}
\label{conclusions}

In this article we have studied how we can apply a cutting operation akin to the one used in the hexagon proposal to a cylindrical worldsheet. As the cutting operator is well defined after being applied once, we can define a square ``form factor'' through a consistency condition coming from subsequent applications of the cutting procedure. We have proven that, applying this construction to an integrable $SU(2)$ model, the consistency condition retrieves the Gaudin determinant. We have also computed off-shell norm of Bethe states for the $SU(2)$ spin chain for the cases of a few excitations.

There exists several possible generalization of the work presented here. The most immediate are the generalizations to the full $PSU(2,2|4)$ symmetry and to general values of the 't Hooft coupling. There also exist two computations that provide relevant checks for the validity of the square and its consistency condition. The first computation we propose is to calculate higher loops and check the proposal for the ``all-loops normalization" presented in eq. (17) of \cite{TailoringIV}, which amount to replace the Heisenberg S-matrix by the all-loops S-matrix and a modification of the $f^{\neq} (\{ u \},\{ u\})$ prefactor. The results obtained here can easily be generalized to inhomogeneous spin chains, so we can use either the $\theta$-morphism \cite{TailoringIV} or the inhomogeneous version of the BDS spin chain \cite{InhomogeneousBDS} to check it. A second computation is to study the extremal limit of the hexagon form factor. In this limit one of the physical sides of the hexagon vanishes, as it has zero length, effectively transforming it into a square. \footnote{The author would like to thank Ivan Kostov and Thiago Fleury for proposing this method.} These computations will be addressed in future works.

\begin{figure}[t]
\begin{center}
\begin{tikzpicture}[scale=1.7]
\draw [densely dashed] (-1/2,1.73205/2) -- (0,0);
\draw (0, 0) -- (1,0);
\draw [densely dashed] (1,0) -- (3/2,1.73205/2);
\draw (3/2,1.73205/2) -- (1,1.73205);
\draw [densely dashed] (1,1.73205) -- (0,1.73205);
\draw (0,1.73205) --(-1/2,1.73205/2);

\draw [fill] (2/3,0) circle [radius=0.04];
\draw [fill] (4/3,4/6*1.73205) circle [radius=0.04];
\draw [thick] (2/3,0) -- (4/3,4/6*1.73205);

\draw [ultra thick] (1.65,1.73205*0.4) -- (2.05,1.73205*0.4);
\draw [ultra thick] (1.65,1.73205*0.6) -- (2.05,1.73205*0.6);

\draw [densely dashed] (2.7-1/2,1.73205/2) -- (2.7+0,0);
\draw (2.7+0, 0) -- (2.7+1,0);
\draw [densely dashed] (2.7+1,0) -- (2.7+3/2,1.73205/2);
\draw (2.7+3/2,1.73205/2) -- (2.7+1,1.73205);
\draw [densely dashed] (2.7+1,1.73205) -- (2.7+0,1.73205);
\draw (2.7+0,1.73205) --(2.7-1/2,1.73205/2);
\draw [densely dashed] (2.7+1.1,-0.0866) -- (2.7+1.6,1.73205/2-0.0866);
\draw (2.7+1.6,1.73205/2-0.0866) -- (2.7+1.6+1.73205/2,1.73205/2-0.5866);
\draw [densely dashed] (2.7+1.6+1.73205/2,1.73205/2-0.5866) -- (2.7+1.1+1.73205/2,-0.5866);
\draw (2.7+1.1+1.73205/2,-0.5866) -- (2.7+1.1,-0.0866);

\draw [fill] (2.7+2/3,0) circle [radius=0.04];
\draw [fill] (2.7+4/3,4/6*1.73205) circle [radius=0.04];
\draw [thick] (2.7+2/3,0) -- (2.7+4/3,4/6*1.73205);

\draw [ultra thick] (5.166,1.73205*0.5) -- (5.566,1.73205*0.5);
\draw [ultra thick] (5.366,1.73205*0.5-0.2) -- (5.366,1.73205*0.5+0.2);

\draw [densely dashed] (6.2-1/2,1.73205/2) -- (6.2+0,0);
\draw (6.2+0, 0) -- (6.2+1,0);
\draw [densely dashed] (6.2+1,0) -- (6.2+3/2,1.73205/2);
\draw (6.2+3/2,1.73205/2) -- (6.2+1,1.73205);
\draw [densely dashed] (6.2+1,1.73205) -- (6.2+0,1.73205);
\draw (6.2+0,1.73205) --(6.2-1/2,1.73205/2);
\draw [densely dashed] (6.2+1.1,-0.0866) -- (6.2+1.6,1.73205/2-0.0866);
\draw (6.2+1.6,1.73205/2-0.0866) -- (6.2+1.6+1.73205/2,1.73205/2-0.5866);
\draw [densely dashed] (6.2+1.6+1.73205/2,1.73205/2-0.5866) -- (6.2+1.1+1.73205/2,-0.5866);
\draw (6.2+1.1+1.73205/2,-0.5866) -- (6.2+1.1,-0.0866);

\draw [fill] (6.2+1.1+1.73205/4,-0.25-0.0866) circle [radius=0.04];
\draw [fill] (6.2+1.6+1.73205/4,1.73205/2-0.25-0.0866) circle [radius=0.04];
\draw [thick] (6.2+1.1+1.73205/4,-0.25-0.0866) -- (6.2+1.6+1.73205/4,1.73205/2-0.25-0.0866);
\end{tikzpicture}
\caption{First non-trivial asymptotic consistency condition involving a hexagon and a square.} \label{hexagontwo}
\end{center}
\end{figure}
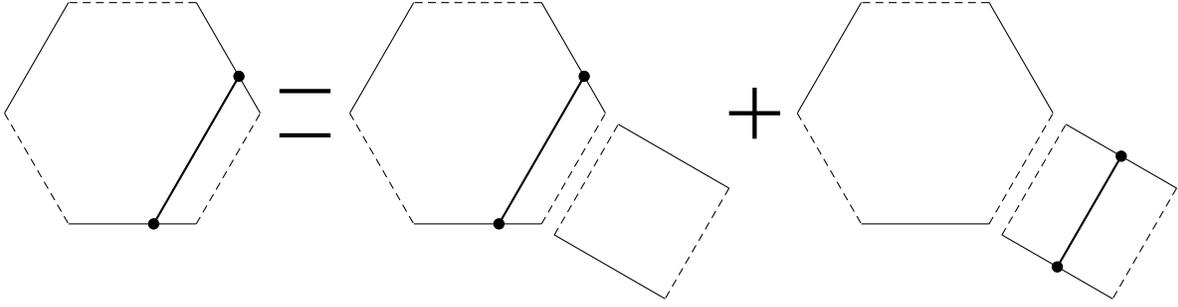

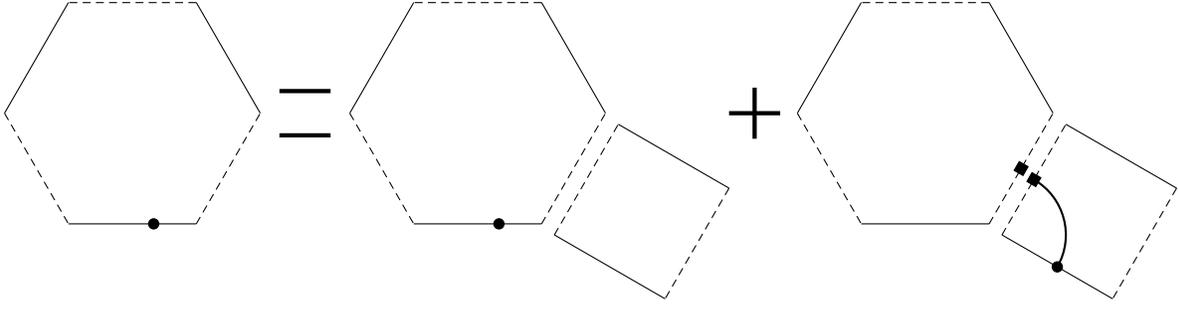
\begin{figure}[t]
\begin{center}
\begin{tikzpicture}[scale=1.7]
\draw [densely dashed] (-1/2,1.73205/2) -- (0,0);
\draw (0, 0) -- (1,0);
\draw [densely dashed] (1,0) -- (3/2,1.73205/2);
\draw (3/2,1.73205/2) -- (1,1.73205);
\draw [densely dashed] (1,1.73205) -- (0,1.73205);
\draw (0,1.73205) --(-1/2,1.73205/2);

\draw [fill] (2/3,0) circle [radius=0.04];

\draw [ultra thick] (1.65,1.73205*0.4) -- (2.05,1.73205*0.4);
\draw [ultra thick] (1.65,1.73205*0.6) -- (2.05,1.73205*0.6);

\draw [densely dashed] (2.7-1/2,1.73205/2) -- (2.7+0,0);
\draw (2.7+0, 0) -- (2.7+1,0);
\draw [densely dashed] (2.7+1,0) -- (2.7+3/2,1.73205/2);
\draw (2.7+3/2,1.73205/2) -- (2.7+1,1.73205);
\draw [densely dashed] (2.7+1,1.73205) -- (2.7+0,1.73205);
\draw (2.7+0,1.73205) --(2.7-1/2,1.73205/2);
\draw [densely dashed] (2.7+1.1,-0.0866) -- (2.7+1.6,1.73205/2-0.0866);
\draw (2.7+1.6,1.73205/2-0.0866) -- (2.7+1.6+1.73205/2,1.73205/2-0.5866);
\draw [densely dashed] (2.7+1.6+1.73205/2,1.73205/2-0.5866) -- (2.7+1.1+1.73205/2,-0.5866);
\draw (2.7+1.1+1.73205/2,-0.5866) -- (2.7+1.1,-0.0866);

\draw [fill] (2.7+2/3,0) circle [radius=0.04];

\draw [ultra thick] (5.166,1.73205*0.5) -- (5.566,1.73205*0.5);
\draw [ultra thick] (5.366,1.73205*0.5-0.2) -- (5.366,1.73205*0.5+0.2);

\draw [densely dashed] (6.2-1/2,1.73205/2) -- (6.2+0,0);
\draw (6.2+0, 0) -- (6.2+1,0);
\draw [densely dashed] (6.2+1,0) -- (6.2+3/2,1.73205/2);
\draw (6.2+3/2,1.73205/2) -- (6.2+1,1.73205);
\draw [densely dashed] (6.2+1,1.73205) -- (6.2+0,1.73205);
\draw (6.2+0,1.73205) --(6.2-1/2,1.73205/2);
\draw [densely dashed] (6.2+1.1,-0.0866) -- (6.2+1.6,1.73205/2-0.0866);
\draw (6.2+1.6,1.73205/2-0.0866) -- (6.2+1.6+1.73205/2,1.73205/2-0.5866);
\draw [densely dashed] (6.2+1.6+1.73205/2,1.73205/2-0.5866) -- (6.2+1.1+1.73205/2,-0.5866);
\draw (6.2+1.1+1.73205/2,-0.5866) -- (6.2+1.1,-0.0866);

\draw [fill,rotate around={-30:(6.2+1+1/4,1.73205/4)}] (6.2+1+1/4-0.04,1.73205/4-0.04) rectangle (6.2+1+1/4+0.04,1.73205/4+0.04);
\draw [fill,rotate around={-30:(6.2+1+1/4+0.1,1.73205/4-0.0866)}] (6.2+1+1/4+0.1-0.04,1.73205/4-0.0866-0.04) rectangle (6.2+1+1/4+0.1+0.04,1.73205/4-0.0866+0.04);
\draw [fill] (6.2+1.1+1.73205/4,-0.25-0.0866) circle [radius=0.04];
\draw[thick] (6.2+1+1/4+0.1,1.73205/4-0.0866) to [out=-30,in=60] (6.2+1.1+1.73205/4,-0.25-0.0866);
\end{tikzpicture}
\caption{First non-trivial consistency condition involving a hexagon and a square if we include mirror contributions.} \label{hexagonmirror}
\end{center}
\end{figure}

However, the most interesting line of work consists in integrating this square proposal with the hexagon proposal. In principle, we expect that the hexagon form factor follows a similar consistency condition to the one proposed here for the square, which in the asymptotic regime should be given by
\begin{multline}
	\mathcal{H} (\{ u \}, \{ v\} , \{ w \} )_{L_1,L_2,L_3}= \\
	=\sum_{\substack{\alpha \cup \bar{\alpha}= \{ u \} \\ \beta \cup \bar{\beta}= \{ v \}}} w (\{ \alpha \} , \{ \bar{\alpha} \} ,\{ \beta \} , \{ \bar{\beta} \} ) \mathcal{H}_{l_1,l_2,L_3} ( \{ \alpha \} , \{ \beta \} , \{ w \} ) S_{l} ( \{ \bar{\alpha} \} \rightarrow \{ \bar{\beta} \} ) \ ,
\end{multline} 
where $l_1+l=L_1$ and $l_2+l=L_2$, and similarly with other pairs of physical sides of the hexagon. Figure~\ref{hexagontwo} shows a pictorial representation of the consistency equation for the hexagon with two excitation. If we decide to include mirror contributions, the first non-trivial consistency condition we can write involves instead the hexagon with one excitation, as we can see in figure~\ref{hexagonmirror}. The problem regarding this consistency condition is that the square constructed here has been defined using a different operator algebra than the one used for the hexagon proposal. Here we have constructed the breaking factors from the commutation relations of the $B$ operator from the Algebraic Bethe Ansatz, while excitations in the hexagon proposal behaves according to a Zamolodchikov-Faddeev algebra. \footnote{For a construction of the hexagon vertex using ZF operators, see section 7.3 of \cite{thesis}.} This implies that the weights associated to both constructions are different, as commented in appendix~\ref{BFappendix}. Despite that, we expect such consistency condition to hold when both form factors are written using the ZF algebra.

Regarding the square described in \cite{HexagonWL}, we expect it to fulfil a consistency condition similar to the one we found here. We denote it by $S^{(L)}_{KK}$ or $S^{(R)}_{KK}$, depending on which edge carries the contribution from the Wilson loop, to distinguish it from the square we are proposing. With this notation, we expect these two kinds of squares to satisfy the following consistency condition.
\begin{align}
	S^{(L)}_{KK,L} (\u \rightarrow \{ v\})=\sum_{\substack{\alpha \cup \bar{\alpha} =\u \\ \beta \cup \bar{\beta} =\{ v \} }} w (\alpha , \bar{\alpha}, \beta, \bar{\beta}) S^{(L)}_{KK,l_1} (\alpha \rightarrow \beta ) S_{l_2} (\bar{\alpha} \rightarrow \bar{\beta} ) \ , \notag \\
	S^{(R)}_{KK,L} (\u \rightarrow \{ v\})=\sum_{\substack{\alpha \cup \bar{\alpha} =\u \\ \beta \cup \bar{\beta} =\{ v \} }} w (\alpha , \bar{\alpha}, \beta, \bar{\beta}) S_{l_1} (\alpha \rightarrow \beta ) S^{(R)}_{KK,l_2} (\bar{\alpha} \rightarrow \bar{\beta} ) \ ,
\end{align}
plus mirror corrections. In the same way that three squares $S_{KK}$ glued to the hexagon give us the three open string interaction in the semiclassical limit, we expect the gluing of $S^{(L)}_{KK}$ with $S^{(R)}_{KK}$ to be equivalent to the propagation of an open string and the gluing of $S_L$ with itself to be equivalent to the propagation of a closed string.

Although only sketched here, the square could be used as a tool for understanding the gluing procedure. The appeal of the square over the hexagon and octagon form factors, for which some interesting results have already been obtained \cite{Hex2,hexagon2,octagon}, is its simplicity. As sketched in appendix~\ref{mirrorappendix}, the gluing of two squares or the gluing of a square and a hexagon form factor involve a finite amount of terms.

\section*{Acknowledgements}

I would like to thank Yunfeng Jiang, Sergey Frolov and Shota Komatsu for interesting discussions and Ivan Kostov, Didina Serban, Thiago Fleury, Minkyoo Kim, Rafael Hernández and Roberto Ruiz for questions and comments on the manuscript. I thank the Institut de Physique Théorique (IPhT) for the hospitality during my visits.

\appendix

\section{Kirchhoff's matrix-tree theorem and Gaudin determinant} \label{Kirchhoffappendix}

In this appendix we review the Kirchhoff's matrix-tree theorem and the results of applying it to the Gaudin determinant.

Given a graph $G$ with $n$ vertices, we can define three matrices that completely define it: the degree matrix, the adjacency matrix and the Laplacian matrix, being the last one the difference of the two previous one. The degree matrix is a diagonal matrix indicating the number of edges attaches to each vertex. The adjacency matrix is a square matrix such that its element $a_{ij}$ with $i\neq j$ is one if there exist an edge fromn vertex $i$ to vertex $j$ and zero if there is not while its elements $a_{ii}$ are equal to zero as we are not going to consider graphs with loops (edges that start and end in the same vertex). If there is a loop in vertex $i$, then we have to set $a_{ii}=2$ instead.

To state the Kirchhoff's theorem we have first to define what a spanning tree is. On the one hand, a graph $H$ with $n$ vertices is said to be a tree if there is a unique path between any two vertices. This condition is equivalent to $H$ being connected and having $n-1$ edges. The connectivity condition can be replaced with the condition of having no cycles without adding any further restriction. A directed tree is defined as a directed acyclic graph whose underlying undirected graph is a tree. A subclass of directed trees are rooted trees, which are defined as directed trees with a natural orientation away from a vertex called root. A forest graph is just an acyclic graph, so it can be understood as a graph consisting of one or more tree graphs. On the other hand, a subgraph $H$ of a graph $G$ is said to be a spanning subgraph if both have the same vertex set. Finally, a spanning rooted tree is a spanning subgraph with is also a rooted tree.

\emph{Kirchhoff's matrix-tree theorem} states that the number of spanning trees contained in a graph $G$ is given by the determinant of any first minors of its Laplacian matrix with respect to diagonal elements, i.e., the Laplacian matrix with row and column $j$ removed for any $j$. This theorem can be generalized to higher degree minors or more complex matrices, we refer to \cite{Tree} and references within for a more comprehensive explanation.

We are interested in the Gaudin determinant, which is the determinant of a matrix of the form
\begin{equation}
	\Phi_{ab}= D_a \delta_{ab} -K_{ab} \ ,
\end{equation}
where $D_a=L \partial_{u_a} p(u_a)$ and $iK_{a,b}=\partial_{u_a} \log S(u_a ,u_b)$. The determinant of this matrix can be expressed as
\begin{equation}
	\det \Phi= \sum_{\text{Forests}} \prod_{\text{roots}} D_a \prod_{\text{edges}} K_{ab} \ . \label{treegaudin}
\end{equation}
The only dependence of the determinant on the length of the spin chain is codified in the terms $D_a$, so the highest power on the length comes from the case of the forest made of isolated vertices. Furthermore, if we break the length $L$ into $l_1+l_2$, we have to add an extra sum over trees having either $l_1 p'$ or $l_2 p'$ as root factor.

\section{More on breaking factors} \label{BFappendix}
In this section we sketch some results about breaking factors beyond rank one. We will also comment on how to modify the construction for operators obeying the Zamolodchikov-Faddeev algebra.

Let us start with the breaking factor for an $SU(N)$ spin chain with R-matrix
\begin{multline}
	R_{1,2} (\lambda) =f (u,v) \sum_{a=1}^N E_{aa} \otimes E_{aa} +\sum_{1\leq a\neq b \leq N} E_{aa} \otimes E_{bb}\\
	+ g(u,v) \sum_{1\leq a\neq b \leq N} (\delta_{a,b+1}+\delta_{a,b-1}) E_{ab} \otimes E_{ba} \ .
\end{multline}
In this case the RTT relation (\ref{RTT}) can be expressed in the two following ways \footnote{The counterpart of these equations for the trigonometric case are more complex and will not be treated here. Their explicit expression can be found for example in eq. (2.4) of \cite{composite}.}
\begin{align}
	\left[ T_{ij} (u) , T_{kl} (v) \right] &= g(u,v) \big( T_{il} (u) T_{kj} (v) - T_{il} (v) T_{kj} (u) \big) \ , \notag \\
	\left[ T_{ij} (u) , T_{kl} (v) \right] &= g(u,v) \big( T_{kj} (v) T_{il} (u) - T_{kj} (u) T_{il} (v) \big) \ .
\end{align}
Writing the monodromy matrix as the product of two monodromy matrices and equating the entries of the auxiliary space we get
\begin{equation}
	T_{ij}^{(L)}=\sum_k T_{ik}^{(l_1)} \otimes T_{kj}^{(l_2)} \ .
\end{equation}
It is more convenient to divide the sum into five different sums depending on if $k$ is between $i$ and $j$, equal to one of them, or greater or lower than both. To clarify the following computations we are going to write $T_{ii}=A_i$, $T_{ij}=B_{ij}$ if $i<j$ and $T_{ij}=C_{ij}$ if $i>j$. To compute the breaking factor we have to compute the breaking of $B_{ij} (u) B_{kl} (v)$ and $C_{kl} (v) C_{ij} (u)$. In this appendix we are only going to examine the cases ``$k=i$ and $l=j$'' and ``$j=i+1$ and $l=k+1$'' for $B$ operators due to the length of the results. We have to take into account that $B$ operators only commute when both indices are equal.

The breaking factor in the first case is extracted from the following expression
\begin{align}
	&B_{ij} (v) B_{ij} (u) = A_i (v) A_i (u) \otimes B_{ij} (v) B_{ij} (u) + B_{ij} (v) B_{ij} (u) \otimes A_j (v) A_j (u) \\
	&+ f(v,u) B_{ij} (v) A_i (u) \otimes B_{ij} (u) A_j (v) + f(u,v) B_{ij} (u) A_i (v) \otimes B_{ij} (v) A_j (u) \notag \\
	&+ \sum_{k=i+1}^{j-1} \left[ f(u,v) B_{ik} (u) A_i (v) \otimes B_{ij} (v) B_{kj} (u) + f(v,u) B_{ik} (v) A_i (u) \otimes B_{ij} (u) B_{kj} (v) \right. \notag \\
	&+ \left. f(v,u) B_{ij} (v) B_{ik} (u) \otimes B_{kj} (u) A_j (v) + f(u,v) B_{ij} (u) B_{ik} (v) \otimes B_{kj} (u) A_j (v)  \right] \ . \notag
\end{align}
It is worth noticing that the first three contributions are equal to the contributions we find for the $SU(2)$ case. In fact, for operators with $j=i+1$ the sum on $k$ vanishes, and we get the same result as in the $SU(2)$ case.

Let us move to the breaking factor of $B_{i,i+1} (u) B_{j,j+1} (v)$. For simplicity, we are going to assume that $i<j$, but a similar computation can be done in the converse case.
\begin{align}
	&B_{i,i+1} (v) B_{j,j+1} (u) = A_i (v) A_j (u) \otimes B_{i,i+1} (v) B_{j,j+1} (u) + B_{i,i+1} (v) B_{j,j+1} (u)  \otimes A_{i+1} (v) A_{j+1} (u) \notag \\
	&+ f(v,u) B_{i,i+1} (v) A_j (u) \otimes B_{j,j+1} (u) A_{i+1} (v) +f(u,v) B_{j,j+1} (u) A_i (v) \otimes B_{i,i+1} (v) A_{j+1} (u) \notag \\
	&+ \delta_{i+1,j} g(v,u) B_{i,i+2} (v) A_{i+1} (u) \otimes \left[ A_{i+1} (u) A_{i+2} (v) - A_{i+1} (v) A_{i+2} (u) \right] \ .
\end{align}
We should pay attention to the appearance of the composite operator $B_{i,i+2}$ in the last equation, making the construction of the weight factors (now weight matrices) more difficult.

These explicit computations show that the construction of weights is more complex in the $SU(N)$ case for $N>2$, as obtaining the weights for higher number of excitations from the weights for two excitations involves summing over internal indices.

To end this section we should comment about ZF operators. Sometimes it is more interesting to work with operators that form a Zamolodchikov-Faddeev algebra instead of using the creation operators from the ABA, as form factors constructed with them fulfil Smirnov's axioms. First of all, we should fix an ordering of the operators, as extra S-matrices appear when they are exchanged. We are going to number the operators in the lower edge from left to right and vice-versa in the upper edge.

As before, computing the weights for this kind of operators requires first to break one single operator into two factors, which is related to the co-product operation. The co-product applied to ZF operators gives
\begin{align}
	\Delta \left[ Z(u) \right] &=Z(u) \otimes \mathbb{I} + H(u) e^{i p(u) l_1} \otimes Z(u) \ , \notag \\ \Delta \left[ Z^\dagger (u) \right] &=\mathbb{I} \otimes Z^\dagger(u) + Z^\dagger (u) \otimes H^{-1}(u) e^{i p(u) l_2}
\end{align}
where $H$ is a well-bred operator in the sense of \cite{ragoucy}. In particular this kind of operators fulfil $H_1 Z_2=S_{12} Z_2 H_1$. Assuming also that $H \ket{0}=\ket{0}$ we obtain the following weights
\begin{align}
	w (\{ \alpha \} , \{ \bar{\alpha} \} ,\{ \beta \} , \{ \bar{\beta} \} ) &= w_Z (\{ \alpha \} , \{ \bar{\alpha} \} ) w_{Z^\dagger} (\{ \beta \} , \{ \bar{\beta} \} ) \label{weightsZF} \\
	w_Z (\{ \alpha \} , \{ \bar{\alpha} \} ) &= e^{i p(\bar{\alpha} )l_1 }S^< (\bar{\alpha},\alpha) \ , \notag \\
	w_{Z^\dagger} (\{ \beta \} , \{ \bar{\beta} \} ) &= e^{i p(\beta )l_2 }S^< (\beta, \bar{\beta}) \ . \notag
\end{align}
Here $S^< (\bar{\alpha},\alpha)$ and $S^< (\beta, \bar{\beta})$ means that we are only keeping the terms $S(u_i,u_j)$ with $i<j$.

A relationship between the square constructed with ZF states and ABA states can be seen already at the level of these weights. If we redefine the ZF square by extracting a factor of $f^{\neq} ({u},{u})$, we can see that the weights become the ones for the ABA square, giving us a simple connection between both squares.

\section{Reconstruction of the spacetime dependence} \label{spacetimedependence}

In this appendix we are going to comment on how to recover the spacetime dependence of the two-point function from two different perspectives. First we are going to make use of the string bits/spin bits construction and later we are going to apply the arguments used in the hexagon proposal.

The string bits/spin bits construction was developed in \cite{vertex1} and \cite{vertex2} and the basic idea is to construct the $\mathcal{N}=4$ SYM operators using the maximal compact subgroup of $U(1) \times SU(2) \times SU(2) \subset PSU(2,2|4)$ and then rotate to the usual representation of the conformal group. Therefore in this language an operator is written as
\begin{align}
	\hat{\mathcal{O}} (x) \left| 0 \right\rangle &= \sum_{\pi\in \mathcal{S}_L} \text{sign} (\pi) \prod_{i=1}^L \left( \hat{O}^{\pi (i)} (x) \left| 0 \right\rangle^{(i)} \right) \ , & \hat{O}^{i} (x) \left| 0 \right\rangle_i= e^{ix P} U O^i \left| 0 \right\rangle ^{(i)} \ ,
\end{align}
where $\text{sign} (\pi)$ only accounts for signs arising from commutating fermions, $\left| 0 \right\rangle ^{(i)} $ represent a vacuum for each bit, $P$ is the momentum operator, $O$ is an operator constructed using such oscillator representation and $U=e^{\frac{\pi}{4} (P_0 - K_0)}$ is an rotation that relates this representation with the usual representation used for classifying fields with respect to the conformal group. This construction allow us to construct a Yangian-invariant spin vertex \cite{vertex2}, which can be understood as the weak-coupling version of string vertex in string field theory and recovers it in the BMN limit \cite{vertex3}.

Regarding the computation of the norm of the two-point function placed at zero and infinity, we can show that it reduces to the scalar product in the Fock space of the oscillators since the inversion of one of the states is equivalent to acting on it with $U^2$
\begin{equation}
	\langle \hat{\mathcal{O}}_j (\infty) \hat{\mathcal{O}}_k (0) \rangle=	\langle \hat{\mathcal{O}}^\dagger_j (0) U^2 \hat{\mathcal{O}}_k (0) \rangle=\frac{1}{L} \sum_{\pi, \sigma\in S_L} \text{sign} (\pi) \text{sign} (\sigma) \prod_{i=1}^L \null_i \left\langle 0 \right| O^{\dagger \pi (i)}_j O^{\sigma (i)}_k \left| 0 \right\rangle_i \ ,
\end{equation}
where we have used that $U^\dagger =U$ and $U^4=1$. If the operators were placed at a finite distance, we would have to commute the exponential of the momentum through the rotation $U$, giving us the tree-level spacetime dependence. The one-loop corrections to the correlation function can be interpreted in this language as the corrections in the commutation relation between the momentum and the $U$ operator (which generate the logarithmic corrections) and from a change in the scalar product of the oscillator (which generate the correction to the mixing matrix). We refer to \cite{vertex1} for the proofs of these statements.

Shifting now to the hexagonalization perspective \cite{Hexagonalization2}, we have computed a square with one operator sit at zero and other at infinity, glued with a second square where we invert the points in which the operators are inserted. The relation between these two frames should be simpler than the twisted translation needed for the hexagon from factor and might impose some constraints to an all-loop construction of the square. The spacetime factor appears as an extra overall dependence in a similar fashion as in the hexagon (see eq. (35), (36) in \cite{Hexagonalization2} and the discussion between them), which is consistent with the way it arises in the string bit/spin bit formalism.

\section{Crossing properties of the square} \label{CrossingSquare}

As we have commented before, we cannot apply the crossing transformation to quantities computed using the Heisenberg spin chain because it corresponds to the vanishing coupling constant limit, and thus the cuts in the rapidity plane necessary for the crossing transformation collapse.

The way to bootstrap a square where we can apply the crossing transformation is to instead construct the consistency condition from the ZF algebra, as in this way we would not be making any assumption on the form of the S-matrix. If we repeat the steps followed in section~\ref{gaudinreconstruction} using the weights (\ref{weightsZF}), we get the same functional form for the square with one excitation
\begin{equation}
	S_L (u\rightarrow v) =\bar{g} (u,v) \left( e^{ip(u) L} -e^{ip(v) L} \right) \ .
\end{equation}
We expect the square from factor to have the following two properties under crossing:
\begin{enumerate}
	\item Moving a excitation clockwise (or anti-clockwise) around the full square should be trivial (up to an S-matrix). We can write this statement as
		\begin{align}
			\bar{g}(u,v)&=\bar{g}(u^c , v)=\bar{g}(v, u^{c} ) S(u^c , v)=\bar{g}(v, u^{2c} ) S(u^c , v) \notag \\
			\bar{g}(u,v)&=\bar{g}(v,u^{-c})=\bar{g}(u^{-c}, v) S(v, u^{-c})=\bar{g}(u^{-2c}, v) S(v, u^{-c}) \ .
		\end{align}
		Where the superindex $c$ means crossing across the mirror side in the clockwise direction and $-c$ means crossing in the counter-clockwise direction.
	\item The crossing of the two excitations is the same as changing the excitations, i.e., $S(u^c \rightarrow v^c )=S(v\rightarrow u)$. This imposes
		\begin{equation}
			\bar{g}(u^c , v^c)=-\bar{g}(v,u) e^{i [p(u) + p(v)] L}=\bar{g}(u^{-c} , v^{-c}) \ .
		\end{equation}
\end{enumerate}
However we are not going to attempt to solve these equations in this article.

\section{Leading mirror corrections to the empty and one excitation squares}\label{mirrorappendix}

In this appendix we are going to perform computations beyond the asymptotic regime by including mirror excitations. There are two contributions we have ignored: The first one is the obvious contribution to the gluing of the mirror sides of a square to obtain the expression for the scalar product. The second one, more subtle, is the contribution to the consistency condition. We are indeed creating new mirror edges when we cut a square into two, thus we have to consider the appearance of mirror excitations in this relation. Here we will make some comments on corrections of the first kind to in the empty square and corrections of second kind to the square with one excitation. 

First of all, since in this section we are going to introduce excitations in the mirror sides, we need a more general notation for the square. This notation will be $S_L (\{u\} | \{v\} | \{w\} | \{x\} )$, where the different sets of rapidities are ordered starting from the lower side and continuing clockwise.

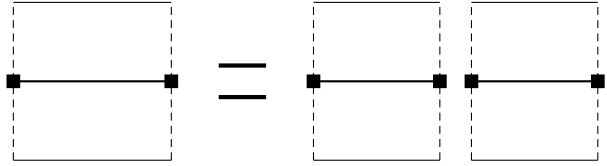
\begin{figure}[t]
\begin{center}
\begin{tikzpicture}[scale=2.1]
\draw [densely dashed] (0,0) --(0,1);
\draw (0, 0) -- (1,0);
\draw (0, 1) -- (1,1);
\draw [densely dashed] (1,0) --(1,1);
\draw [fill] (-0.04,0.5-0.04) rectangle (0.04,0.5+0.04);
\draw [fill] (1-0.04,0.5-0.04) rectangle (1+0.04,0.5+0.04);
\draw [thick] (0,0.5) -- (1,0.5);

\draw [ultra thick] (1.3,0.4) -- (1.6,0.4);
\draw [ultra thick] (1.3,0.6) -- (1.6,0.6);
\draw [densely dashed] (1.9,0) --(1.9,1);
\draw (1.9, 0) -- (2.7,0);
\draw (1.9, 1) -- (2.7,1);
\draw [densely dashed] (2.7,0) --(2.7,1);
\draw [fill] (1.9-0.04,0.5-0.04) rectangle (1.9+0.04,0.5+0.04);
\draw [fill] (2.7-0.04,0.5-0.04) rectangle (2.7+0.04,0.5+0.04);
\draw [thick] (1.9,0.5) -- (2.7,0.5);
\draw [densely dashed] (2.9,0) --(2.9,1);
\draw (2.9, 0) -- (3.7,0);
\draw (2.9, 1) -- (3.7,1);
\draw [densely dashed] (3.7,0) --(3.7,1);
\draw [fill] (2.9-0.04,0.5-0.04) rectangle (2.9+0.04,0.5+0.04);
\draw [fill] (3.7-0.04,0.5-0.04) rectangle (3.7+0.04,0.5+0.04);
\draw [thick] (2.9,0.5) -- (3.7,0.5);
\end{tikzpicture}
\caption{Pictorial representation of the cutting of the first mirror correction to the empty square.} \label{emptymirror}
\end{center}
\end{figure}

We also have to fix our notation regarding the mirror transformation. The rules regarding this transformation can be found, for example, in appendix D of \cite{BKV}. In particular, we are going to define moving an excitation to the next side in the clockwise direction with the superscript $\gamma$, while moving counter-clockwise will be labelled by $-\gamma$. Applying two times this transformation corresponds to a crossing transformation, $u^{2\gamma}=u^c$, which changes the sign of momentum and energy. For simplicity and concreteness we are going to choose the spin chain model with $a(u)=e^{i p(u) L}$ and $d(u)=1$, where we have
\begin{align}
	\tilde{a} (u) &= a( u^{-\gamma})= 1 \ , & \hat{a} (u) &=a (u^{+\gamma}) = e^{-E(u) R} \ , \\
	\tilde{d} (u) &= d( u^{-\gamma})= e^{-E(u) R} \ , & \hat{d} (u) &=d (u^{+\gamma}) = 1 \ .
\end{align}

Let us consider first the case of the first mirror correction to the empty square. The consistency condition, pictorially represented in figure~\ref{emptymirror}, is
\begin{equation}
	S_L (\emptyset | u | \emptyset | w)= \int{\frac{dv}{2\pi} S_{l_1} (\emptyset | u | \emptyset | v) S_{l_2} (\emptyset | v | \emptyset | w) } \ .
\end{equation}
If we substitute here the ansatz $S_L (\emptyset | u | \emptyset | w)=\hat{g}(u,w) [e^{-E(w) R} - e^{-E(u) R}]$, the equation reduces to the sum of the residues at the poles of the function $\hat{g}$. The function $\tilde{g}(u,w)\propto \frac{1}{u-w}$ we obtained for the Gaudin square gives us the correct measure and Boltzmann factor when we take the limit $w\rightarrow u$, but it is not a solution of this equation as the integrand is regular at $v=u$ and $v=w$. On the other hand, this equation can be solved if we use a square obtained from using ZF operators since, according to Smirnov's axioms, it presents the particle-antipartice residue. This fact can be understood as the consistency condition for mirror excitations being sensible to the kind of operators we are using for constructing our squares. In particular, it forces us to use Zamolodchikov-Faddeev operators, as the form factors constructed with them fulfil the Smirnov's axioms.

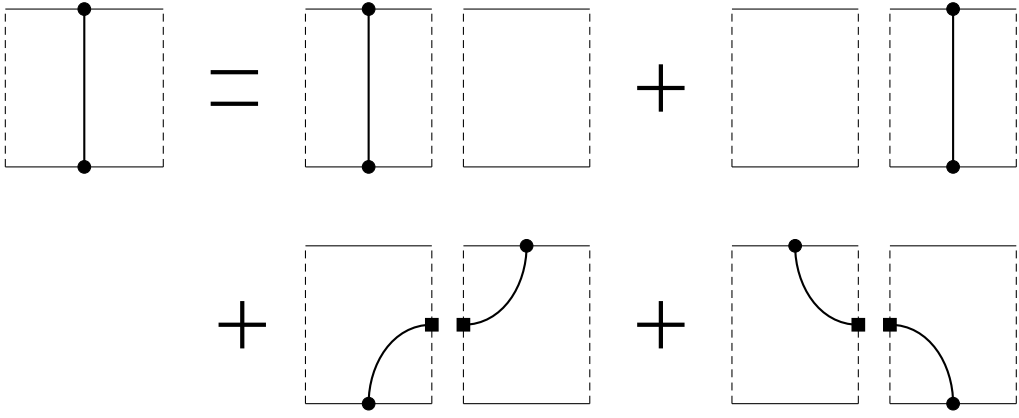
\begin{figure}[t]
\begin{center}
\begin{tikzpicture}[scale=2.1]
\draw [densely dashed] (0,0) --(0,1);
\draw (0, 0) -- (1,0);
\draw (0, 1) -- (1,1);
\draw [densely dashed] (1,0) --(1,1);
\draw [fill] (0.5,0) circle [radius=0.04];
\draw [fill] (0.5,1) circle [radius=0.04];
\draw [thick] (0.5,0) -- (0.5,1);

\draw [ultra thick] (1.3,0.4) -- (1.6,0.4);
\draw [ultra thick] (1.3,0.6) -- (1.6,0.6);
\draw [densely dashed] (1.9,0) --(1.9,1);
\draw (1.9, 0) -- (2.7,0);
\draw (1.9, 1) -- (2.7,1);
\draw [densely dashed] (2.7,0) --(2.7,1);
\draw [fill] (2.3,0) circle [radius=0.04];
\draw [fill] (2.3,1) circle [radius=0.04];
\draw [thick] (2.3,0) -- (2.3,1);
\draw [densely dashed] (2.9,0) --(2.9,1);
\draw (2.9, 0) -- (3.7,0);
\draw (2.9, 1) -- (3.7,1);
\draw [densely dashed] (3.7,0) --(3.7,1);
\draw [ultra thick] (4,0.5) -- (4.3,0.5);
\draw [ultra thick] (4.15,0.35) -- (4.15,0.65);
\draw [densely dashed] (4.6,0) --(4.6,1);
\draw (4.6, 0) -- (5.4,0);
\draw (4.6, 1) -- (5.4,1);
\draw [densely dashed] (5.4,0) --(5.4,1);
\draw [densely dashed] (5.6,0) --(5.6,1);
\draw (5.6, 0) -- (6.4,0);
\draw (5.6, 1) -- (6.4,1);
\draw [densely dashed] (6.4,0) --(6.4,1);
\draw [fill] (6,0) circle [radius=0.04];
\draw [fill] (6,1) circle [radius=0.04];
\draw [thick] (6,0) -- (6,1);

\draw [ultra thick] (1.35,0.5-1.5) -- (1.65,0.5-1.5);
\draw [ultra thick] (1.5,0.35-1.5) -- (1.5,0.65-1.5);

\draw [densely dashed] (1.9,0-1.5) --(1.9,1-1.5);
\draw (1.9, 0-1.5) -- (2.7,0-1.5);
\draw (1.9, 1-1.5) -- (2.7,1-1.5);
\draw [densely dashed] (2.7,0-1.5) --(2.7,1-1.5);
\draw [fill] (2.3,0-1.5) circle [radius=0.04];
\draw [fill] (2.7-0.04,0.5-1.5-0.04) rectangle (2.7+0.04,0.5-1.5+0.04);
\draw [thick] (2.3,0-1.5) to [out=90,in=180] (2.7,0.5-1.5);

\draw [densely dashed] (2.9,0-1.5) --(2.9,1-1.5);
\draw (2.9, 0-1.5) -- (3.7,0-1.5);
\draw (2.9, 1-1.5) -- (3.7,1-1.5);
\draw [densely dashed] (3.7,0-1.5) --(3.7,1-1.5);
\draw [fill] (3.3,1-1.5) circle [radius=0.04];
\draw [fill] (2.9-0.04,0.5-1.5-0.04) rectangle (2.9+0.04,0.5-1.5+0.04);
\draw [thick] (3.3,1-1.5) to [out=-90,in=0] (2.9,0.5-1.5);

\draw [ultra thick] (4,0.5-1.5) -- (4.3,0.5-1.5);
\draw [ultra thick] (4.15,0.35-1.5) -- (4.15,0.65-1.5);

\draw [densely dashed] (4.6,0-1.5) --(4.6,1-1.5);
\draw (4.6, 0-1.5) -- (5.4,0-1.5);
\draw (4.6, 1-1.5) -- (5.4,1-1.5);
\draw [densely dashed] (5.4,0-1.5) --(5.4,1-1.5);
\draw [fill] (5,1-1.5) circle [radius=0.04];
\draw [fill] (5.4-0.04,0.5-1.5-0.04) rectangle (5.4+0.04,0.5-1.5+0.04);
\draw [thick] (5,1-1.5) to [out=-90,in=180] (5.4,0.5-1.5);

\draw [densely dashed] (5.6,0-1.5) --(5.6,1-1.5);
\draw (5.6, 0-1.5) -- (6.4,0-1.5);
\draw (5.6, 1-1.5) -- (6.4,1-1.5);
\draw [densely dashed] (6.4,0-1.5) --(6.4,1-1.5);
\draw [fill] (6,0-1.5) circle [radius=0.04];
\draw [fill] (5.6-0.04,0.5-1.5-0.04) rectangle (5.6+0.04,0.5-1.5+0.04);
\draw [thick] (6,0-1.5) to [out=90,in=0] (5.6,0.5-1.5);
\end{tikzpicture}
\caption{Pictorial representation of the cutting of the square with one excitation, including terms that involve mirror excitations.} \label{onemirror}
\end{center}
\end{figure}

Moving now to the case of one physical excitation, figure~\ref{onemirror} represents the terms appearing in the consistency equation~\ref{bootstrap} when we allow for mirror corrections. We can write the consistency condition in the following way
\begin{align}
	S_L (u|\emptyset |w| \emptyset ) &= S_{l_1} (u|\emptyset |w| \emptyset ) d_{l_2} (u) a_{l_2} (w) +S_{l_2} (u|\emptyset |w| \emptyset ) a_{l_1} (u) d_{l_1} (w) \notag \\
	&+ \int{ \frac{dv}{2\pi} S_{l_1} (u|\emptyset | \emptyset|v ) S_{l_2} (\emptyset | v | w | \emptyset )} d_{l_2} (u) d_{l_1} (w) \notag \\
	&+ \int{ \frac{dv}{2\pi} S_{l_1} (\emptyset |v| \emptyset | w ) S_{l_2} (u | v | \emptyset | \emptyset )} a_{l_1} (u) a_{l_2} (w)  \ .
\end{align}
In contrasts with the gluing of two hexagons, here only two new corrections appear. This is because we have no mirror excitations in the original mirror sides and only two physical excitations to annihilate the set of mirror excitations we add, so we are forced to put at most only one mirror excitation if we want a non-vanishing contribution due to symmetry.

To solve the consistency equation we substitute the ansatz
\begin{equation}
	S_L (u\rightarrow w)= \alpha (u,w) a(u) a(w) + \beta(u,w) a(u) d(w) + \gamma(u,w) d(u) a(w) + \delta(u,w) d(u) d(w) \ ,
\end{equation}
reducing it to
\begin{align*}
	\alpha_L &= \hat{I}_1 \ , & \beta_L &= \beta_{l_2} \ , & \gamma_L &=\gamma_{l_1} \ , & \delta_L &= \tilde{I}_4 \ , \\
	\alpha_{l_1} &= -\hat{I}_3 \ , & \delta_{l_1} &= -\tilde{I}_2 \ , & \alpha_{l_2} &= -\hat{I}_2 \ , & \delta_{l_2} &= -\tilde{I}_3 \ ,
\end{align*}
together with $\beta_{l_1} + \gamma_{l_2}+\tilde{I}_1 + \hat{I}_4=0$. The different $I$'s are given by the following integrals
\begin{align*}
	\tilde{I}_1 &=\int{\frac{dv}{2\pi} \left[ \alpha_{l_1} \alpha_{l_2} [\tilde{a} (v)]^2 + \left( \alpha_{l_1} \gamma_{l_2}  + \beta_{l_1} \alpha_{l_2} \right) \tilde{a} (v) \tilde{d} (v)+ \gamma_{l_1} \gamma_{l_2} [\tilde{d} (v)]^2  \right]} \ , \\
	\tilde{I}_2 &=\int{\frac{dv}{2\pi} \left[ \gamma_{l_1} \alpha_{l_2} [\tilde{a} (v)]^2 + \left( \gamma_{l_1} \gamma_{l_2}  + \delta_{l_1} \alpha_{l_2} \right) \tilde{a} (v) \tilde{d} (v)+ \delta_{l_1} \gamma_{l_2} [\tilde{d} (v)]^2  \right]}  \ , \\
	\tilde{I}_3 &=\int{\frac{dv}{2\pi} \left[ \alpha_{l_1} \beta_{l_2} [\tilde{a} (v)]^2 + \left( \alpha_{l_1} \delta_{l_2}  + \beta_{l_1} \beta_{l_2} \right) \tilde{a} (v) \tilde{d} (v)+ \gamma_{l_1} \delta_{l_2} [\tilde{d} (v)]^2  \right]}  \ , \\
	\tilde{I}_4 &=\int{\frac{dv}{2\pi} \left[ \gamma_{l_1} \beta_{l_2} [\tilde{a} (v)]^2 + \left( \gamma_{l_1} \delta_{l_2}  + \delta_{l_1} \beta_{l_2} \right) \tilde{a} (v) \tilde{d} (v)+ \delta_{l_1} \delta_{l_2} [\tilde{d} (v)]^2  \right]} \ , \\
	\hat{I}_1 &=\int{\frac{dv}{2\pi} \left[ \alpha_{l_1} \alpha_{l_2} [\hat{a} (v)]^2 + \left( \gamma_{l_1} \alpha_{l_2}  + \alpha_{l_1} \beta_{l_2} \right) \hat{a} (v) \hat{d} (v)+ \gamma_{l_1} \beta_{l_2} [\hat{d} (v)]^2  \right]}  \ , \\
	\hat{I}_2 &=\int{\frac{dv}{2\pi} \left[ \beta_{l_1} \alpha_{l_2} [\hat{a} (v)]^2 + \left( \delta_{l_1} \alpha_{l_2}  + \beta_{l_1} \beta_{l_2} \right) \hat{a} (v) \hat{d} (v)+ \delta_{l_1} \beta_{l_2} [\hat{d} (v)]^2  \right]} \ , \\
	\hat{I}_3 &=\int{\frac{dv}{2\pi} \left[ \alpha_{l_1} \gamma_{l_2} [\hat{a} (v)]^2 + \left( \gamma_{l_1} \gamma_{l_2}  + \alpha_{l_1} \delta_{l_2} \right) \hat{a} (v) \hat{d} (v)+ \gamma_{l_1} \delta_{l_2} [\hat{d} (v)]^2  \right]} \ , \\
	\hat{I}_4 &=\int{\frac{dv}{2\pi} \left[ \beta_{l_1} \gamma_{l_2} [\hat{a} (v)]^2 + \left( \delta_{l_1} \gamma_{l_2}  + \beta_{l_1} \delta_{l_2} \right) \hat{a} (v) \hat{d} (v)+ \delta_{l_1} \delta_{l_2} [\hat{d} (v)]^2  \right]} \ .
\end{align*}
The arguments of the functions involved in the tilde integrals are $(u,v^{-\gamma})$ and $(v^{-\gamma},w)$ for the functions with subindexes $l_1$ and $l_2$ respectively, while for the hat integrals the arguments are $(v^{\gamma},w)$ and $(u,v^{\gamma})$ for $l_1$ and $l_2$ respectively. Note that not all these integrals are independent, as the square should be invariant under the interchange of the initial and final physical rapidities. This implies the following relations between the unknown functions of the ansatz
\begin{align}
	\alpha (u,w) &= \alpha (w,u) \ , & \delta (u,w) &=\delta (w,u) \ , & \beta (u,w) &= \gamma (w,u) \ .
\end{align}
Using these properties we can prove some consistency of this ansatz as $\alpha_{l_1}=\hat{I}_2 (u,w) = \hat{I}_3 (w,u)=\alpha_{l_2}$ and similarly for the tilde ones, which implies that both $\alpha$ and $\delta$ are independent of the length. Furthermore, as $I_1$'s and $I_4$'s transform into themselves, we can write function $\beta$ and $\gamma$ as
\begin{equation}
	\beta (u,w)-\frac{\tilde{I}_1 +\hat{I}_4}{2} = \gamma (u,w)+\frac{\tilde{I}_1 +\hat{I}_4}{2} =\tilde{g}(u,v) \ ,
\end{equation}
where $\tilde{g}$ is an arbitrary function with the property $\tilde{g}(u,v)=-\tilde{g}(v,u)$. Using these properties the problem reduces to a system of coupled Fredholm integral equations of second kind, which can be solved as a Liouville-Neumann series. However, we need an initial condition to apply this method. As we have seen for the case of the mirror excitation in the empty square, the initial condition we have to use is the square computed using ZF operators.


\end{document}